\documentclass[aps,prx,reprint,showpacs,floatfix,superscriptaddress]{revtex4-1}
\usepackage{xcolor}
\usepackage{microtype}
\usepackage[bookmarks=true,
bookmarksnumbered=false, 
bookmarksopen=false, 
colorlinks=true,
linkcolor=blue]{hyperref}
\definecolor{webblue}{rgb}{0, 0, 0.5} 
\hypersetup{colorlinks=true,urlcolor=blue,citecolor=blue}
\usepackage{graphicx}
\usepackage{subfigure}
\usepackage{url}
\usepackage{hyperref}
\usepackage{inconsolata}
\usepackage{amsmath}
\usepackage[capitalize]{cleveref}
\usepackage{braket}
\usepackage{outlines}
\usepackage{enumitem}
\usepackage{soul}
\usepackage{color}
\usepackage{bm} 

\usepackage[utf8]{inputenc}
\usepackage{siunitx,booktabs}
\usepackage{multirow}
\usepackage{amssymb}

 \usepackage{relsize}

\usepackage{soul}

\usepackage{pgfplots}
\usepackage{ifthen}

\begin{document}

 \title{Artificial Graphene: Unconventional Superconductivity in a Honeycomb Superlattice}

\author{Tommy Li}
\email{tommyli@zedat.fu-berlin.de}
\affiliation{Dahlem Center for Complex Quantum Systems and Fachbereich Physik, Freie Universit\"{a}t Berlin, Arnimallee 14, 14195 Berlin, Germany}
\affiliation{Center for Quantum Devices, Niels Bohr Institute, University of Copenhagen, DK-2100 Copenhagen, Denmark}
\author{Julian Ingham}
\affiliation{Physics Department, Boston University, Commonwealth Avenue, Boston, MA 02215, USA}
\author{Harley D. Scammell}
\affiliation{Department of Physics, Harvard University, Cambridge, MA 02138, USA}

\date{\today}

\begin{abstract}
Artificial lattices have served as a platform to study the physics of unconventional superconductivity. We study semiconductor artificial graphene -- a honeycomb superlattice imposed on a semiconductor heterostructure -- which hosts the Dirac physics of graphene but with a tunable periodic potential strength and lattice spacing, allowing control of the strength  of the electron-electron interactions. We demonstrate a new mechanism for superconductivity due to repulsive interactions which requires a strong lattice potential and a minimum doping away from the Dirac points. The mechanism relies on the Berry phase of the emergent Dirac fermions, which causes oppositely moving electron pairs near the Dirac points to interfere destructively, reducing the Coulomb repulsion and thereby giving rise to an effective attraction. The attractive component of the interaction is enhanced by a novel antiscreening effect which, in turn, increases with doping; as a result there is a minimum doping beyond which superconducting order generically ensues. The dominant superconducting state exhibits a spatially modulated gap with chiral $p$-wave symmetry. Microscopic calculations suggest that the possible critical temperatures are large relative to the low carrier densities, for a range of experimentally realistic parameters.
\end{abstract}

\maketitle

\section{Introduction}
\label{intro}

Two dimensional semiconductor systems have provided striking manifestations of both the quantum behavior of single electrons and a variety of paradigmatic interacting states of matter \cite{SemiC}. Over several decades, experimental technology has advanced to allow remarkable control and tunability over these systems, and access to a variety of fundamental physical effects. Designer superlattices such as artificial graphene (AG) -- a semiconductor heterostructure patterned with a  honeycomb lattice potential -- seek to combine the novel physics of materials like graphene with the high degree of control in semiconductor devices  \cite{Park2009,Gibertini2009,Singha2011,AG,AG2,Sushkov2013,Tkachenko2015,Li2016,Li2017, Scammell2019,Soibel1996}. As in conventional graphene, the periodic potential in AG gives rise to a pair of band crossings near which the single--electron dynamics may be described by a $2+1$ dimensional Dirac fermionic theory with emergent relativistic invariance. Motivated by substantial recent improvements in the quality of these superlattices \cite{AG,AG2}, we propose that such a system is capable of supporting a new type of unconventional superconductivity, across a range of experimentally achievable parameters.

Our mechanism relies on the Berry phase associated with the emergent Dirac fermions of the superlattice, which provides the attraction between electrons via a novel interference effect, {that} promotes Cooper pairing as a way to lower the energy cost of Coulomb repulsion. We find that this mechanism is effective when the atomic orbitals of the superlattice become localized, unlike in graphene, a regime which can be reached by narrowing or deepening the minima of the superlattice potential. We do not rely on nesting, van Hove singularities,  the Kohn--Luttinger effect, or spin fluctuations near a magnetically ordered state. Rather, the pairing is mediated by fluctuations of an emergent pseudospin degree of freedom. While we expect our mechanism to be a universal feature of interacting Dirac systems, including twisted bilayer graphene and similar twisted layered systems, we suggest that in semiconductor superlattices, the pairing interaction may be sufficiently enhanced by device engineering to yield values of $T_c$ significantly higher than those either proposed theoretically or observed experimentally in honeycomb lattice materials. We find the dominant instability to occur in the $p+ip$ channel with pairing within the same valley, resulting in a time reversal invariant, spin triplet  gap with finite quasimomentum, furnishing an example of the rare Fulde--Ferrell--Larkin--Ovchinnikov (FFLO) phase \cite{Fulde1964,Larkin1965,Roy2010,Tsuchiya2016}. 

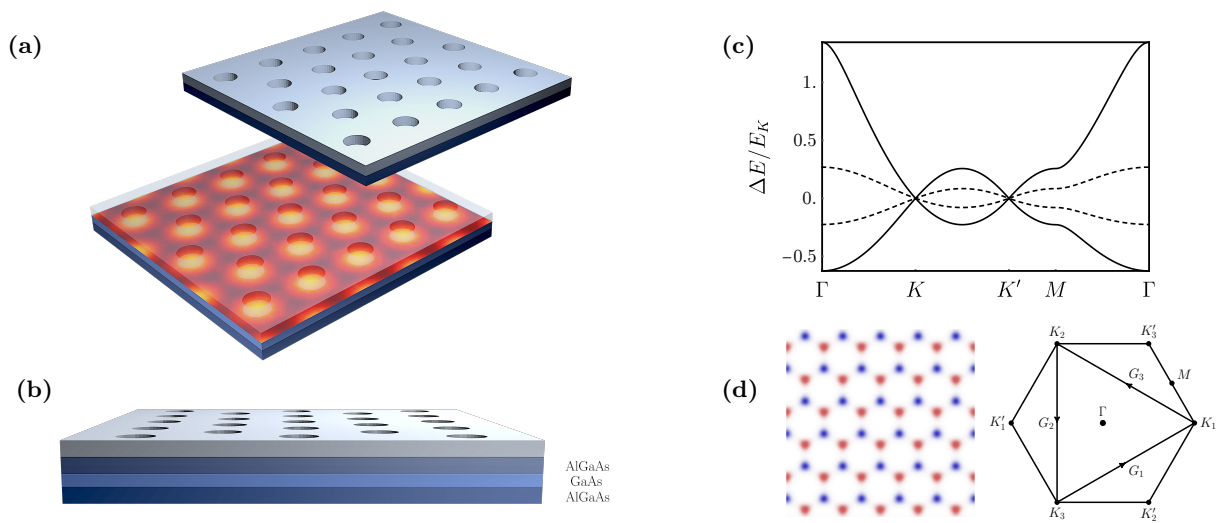
\begin{figure*}[t]

\includegraphics[width=65mm]{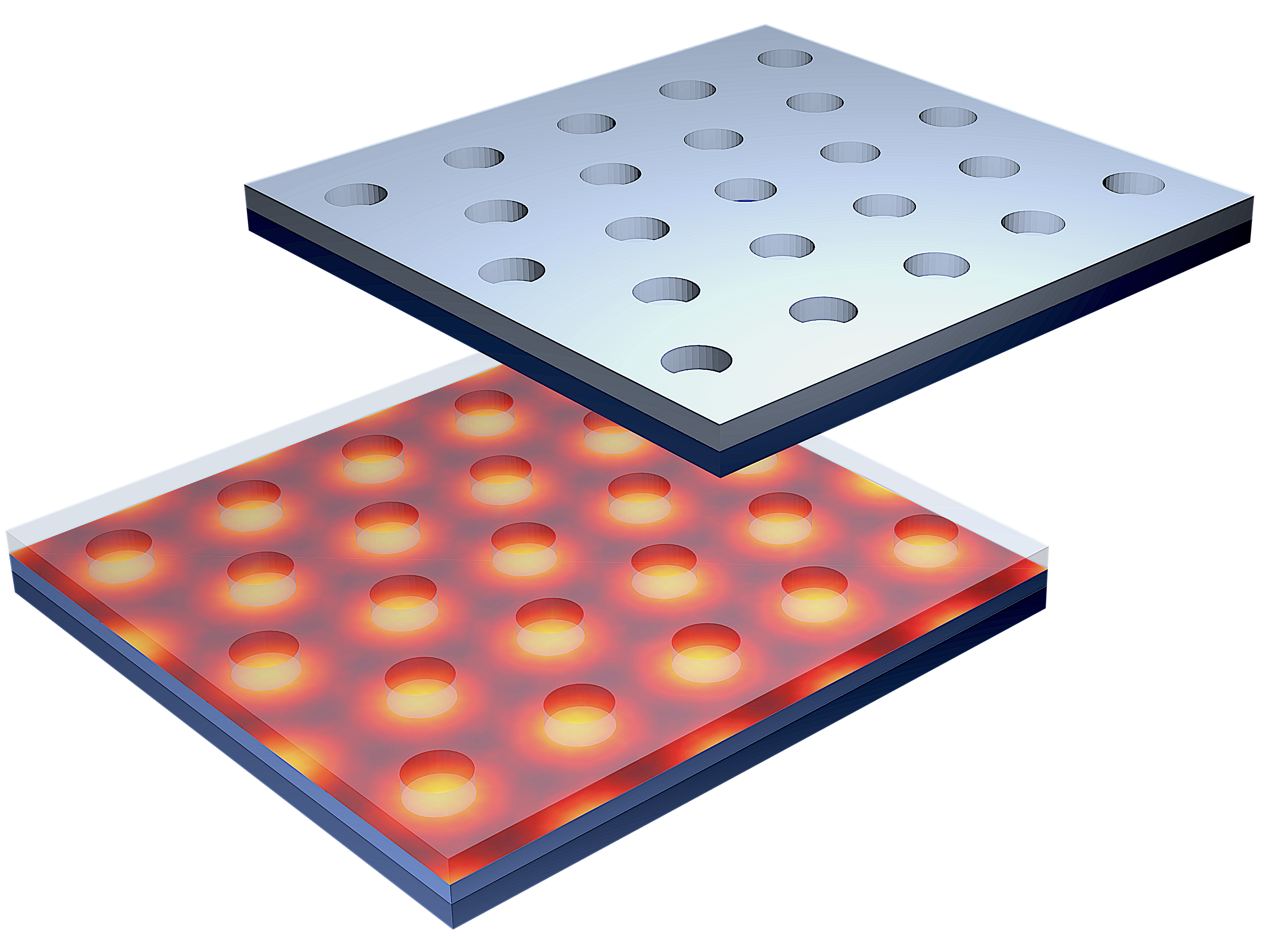} \hspace{2cm}
\raisebox{0.08\height}{\includegraphics[width=55mm]{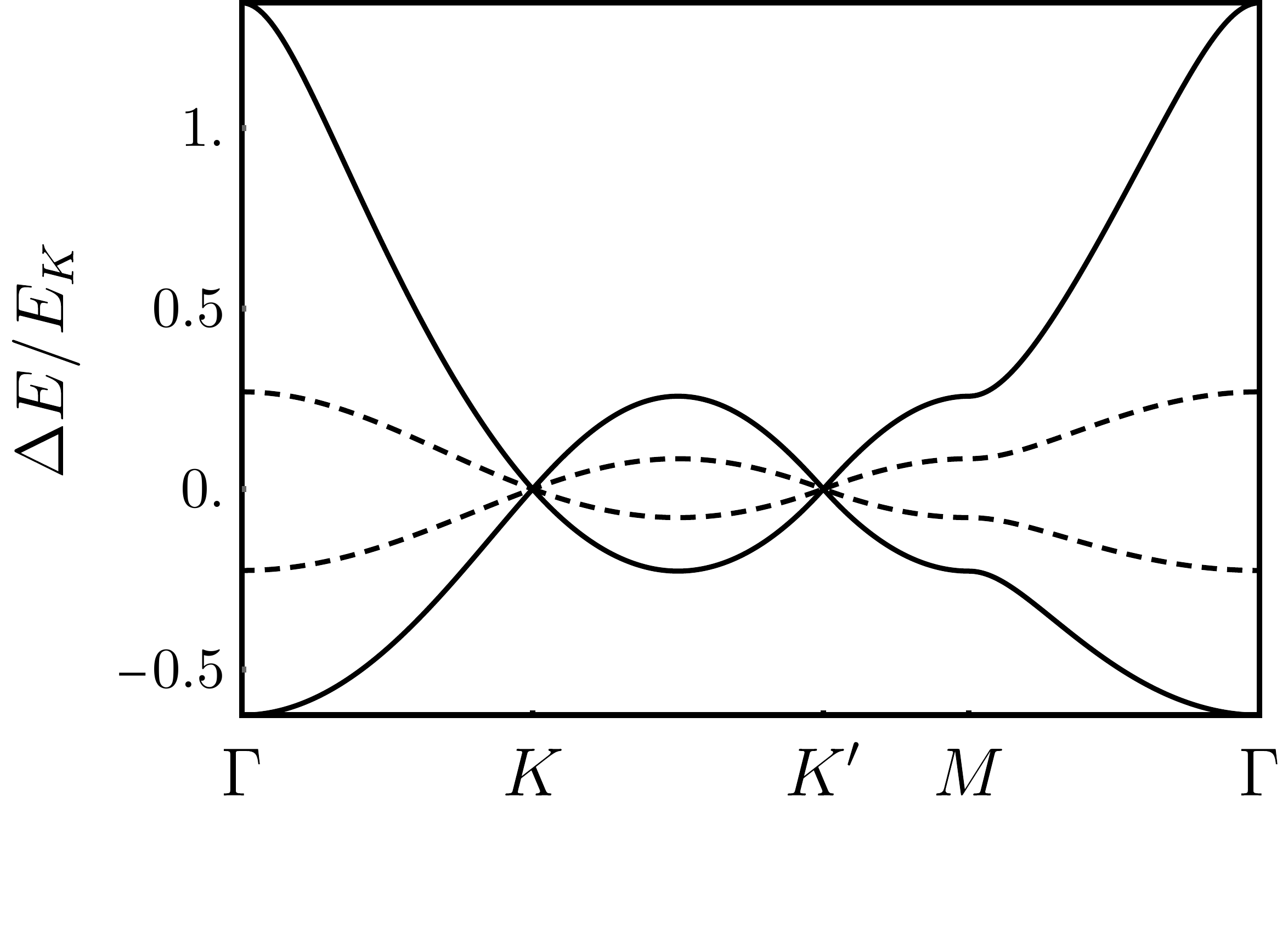}}\vspace{-0.9cm}
\flushleft\hspace{0.8cm}
\raisebox{-0.6\height}{\includegraphics[width=90mm]{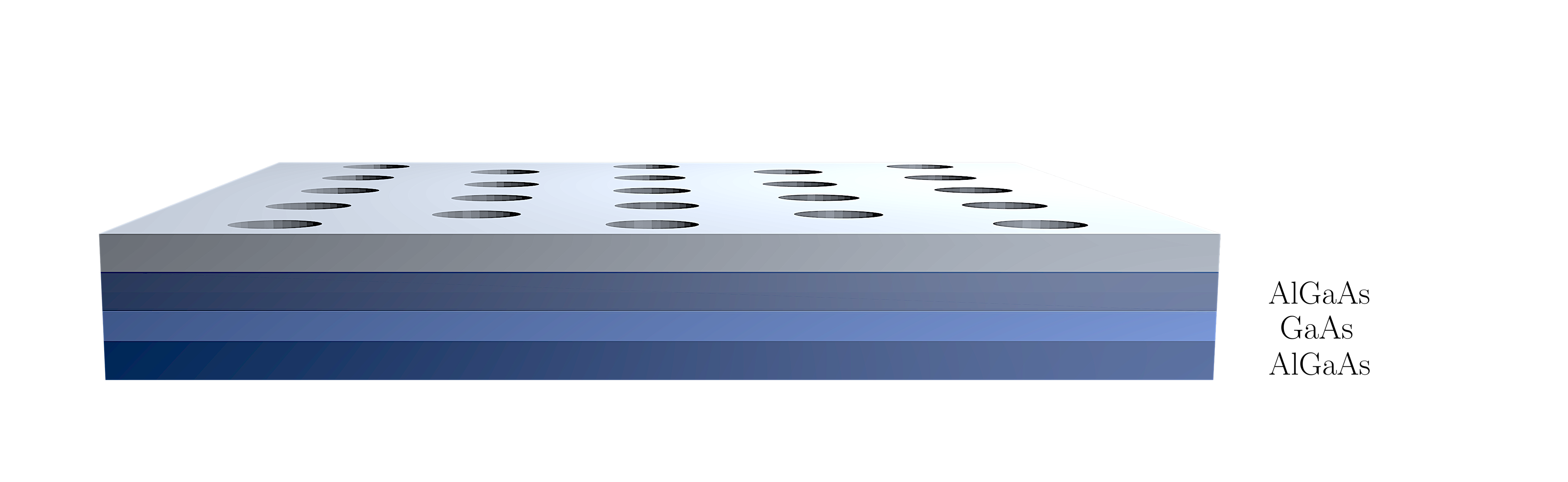}} \hspace{1.0cm}
\raisebox{-0.5\height}{\includegraphics[width=25mm]{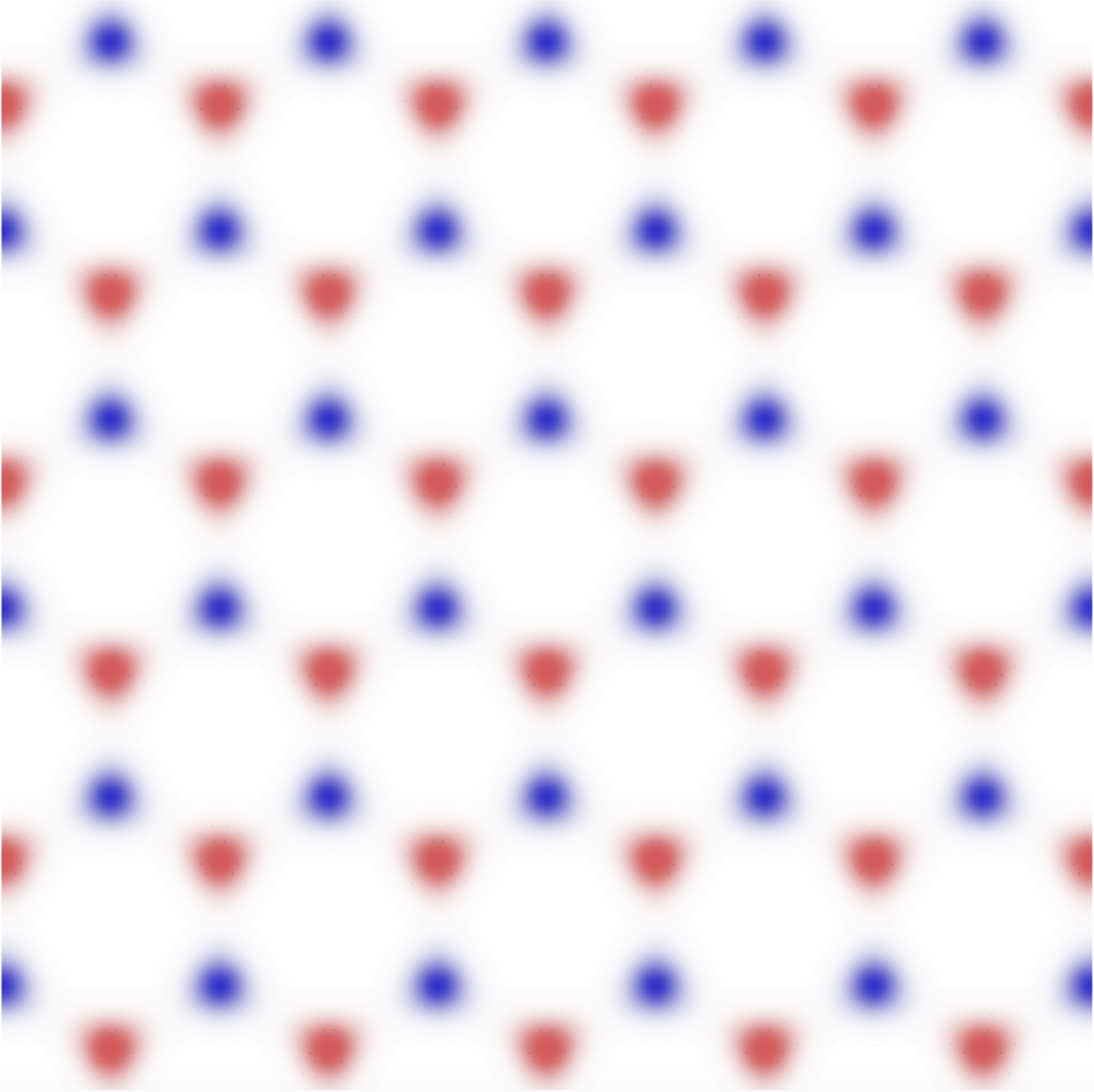}}\hspace{0.1cm}
\raisebox{-0.5\height}{\includegraphics[width=30mm]{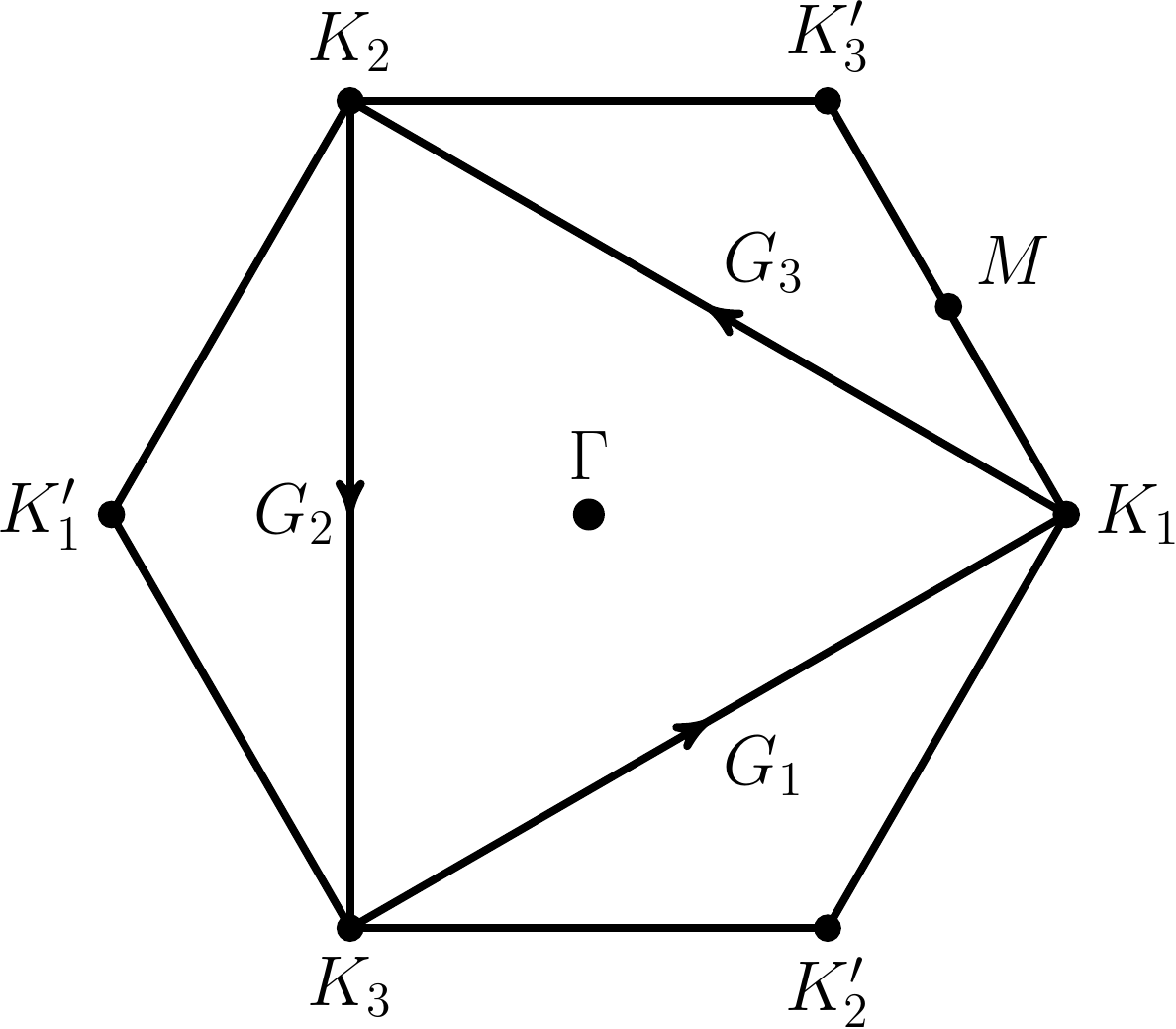}}
\begin{picture}(0,0) 
\put(-460,140){\textbf{(a)}} 
\put(-190,140){\textbf{(c)}} 
\put(-190,10){\textbf{(d)}} 
\put(-460,10){ \textbf{(b)}} 
\end{picture}
\caption{ \textbf{(a)} and \textbf{(b)} Schematic of semiconductor artificial graphene: a 2DEG is confined by a semiconductor heterostructure, chosen for illustration to be AlGaAs--GaAs--AlGaAs. A metallic gate or mask with a triangular antidot array is placed on top of the heterostructure, introducing the hexagonal superlattice potential experienced by the electrons in the 2DEG. \textbf{(c)} The miniband dispersion $E(k)$ in units of $ E_K = K^2/(2m^*)$ where $K = 4\pi/(3L)$ is the valley momentum, plotted for $W = E_K$ (solid), $8E_K$ (dashed). As $W/E_K$ increases and the Fermi velocity decreases, the interactions $V_{00;00}$ and $V_{zz;00}$ in (\ref{bareint}) dominate. \textbf{(d, left)} Density plot of the wavefunctions of (\ref{hamiltonian}), showing pseudospin up (blue) and down (red). \textbf{(d, right)} The Brillouin zone corresponding to the periodic potential (\ref{potential}).  \label{f:chirpa}}
\end{figure*}
Since the single--particle physics in semiconductor heterostructures can be engineered with great control, the known structure of the wavefunctions as well as the possibility of tuning the system into a weakly interacting regime provides a quantitatively reliable starting point for a perturbative examination of interaction effects, in stark contrast to theories of superconductivity in  strongly correlated systems. We are able to predict critical temperatures $T_c\approx 20$ K in InAs quantum wells with lattice spacing $L =$ 10 nm, and $T_c\approx 10$ K for GaAs. The ratio of the critical temperature to the electron density suggests that the electron pairing strength ranks among cuprates, iron pnictides, and twisted bilayer graphene \cite{Cao2018} -- offering a new class of systems to study the elusive physics of strongly bound Cooper pairs.

The paper is structured as follows. In Section \ref{model} we derive the theory describing the physics of AG. In Section \ref{mech} we explain the pseudospin mechanism for superconductivity. In Section \ref{gap} we present the solution to the gap equation. Section \ref{disc} will discuss our results.

\section{Theoretical Model}
\label{model}
The single--particle physics of AG is described by a two dimensional electron gas (2DEG) in the presence of an electrostatic superlattice potential. There are several existing approaches to implementing the superlattice, including patterning the upper layer of a semiconductor heterostructure \cite{AG,AG2} or depositing a metallic top-gate using standard lithographic techniques \cite{Soibel1996}. Accounting for the Coulomb interaction, we may model the system by the second quantized Hamiltonian
\begin{align}
\label{hamiltonian}
\notag H = \int{ \Psi^\dagger(\bm{r})\left[ \frac{p^2}{2m^*} +\mathcal{U}(\bm{r})\right] \Psi(\bm{r}) d^2 \bm{r}} \\
+ \frac{e^2}{2\epsilon_r}\int{ \frac{ \Psi^\dagger(\bm{r}') \Psi(\bm{r}') \Psi^\dagger(\bm{r}) \Psi(\bm{r})}{ |\bm{r}- \bm{r}'|} d^2 \bm{r} d^2 \bm{r}'}
\end{align}
where $m^*$ and $\epsilon_r$ are the effective mass and dielectric constant of the semiconductor, $\bm{r} = (x,y)$ is the in--plane coordinate vector, $p$ is the magnitude of the in--plane electron momentum, $\Psi(\bm{r})$ is the spin-$\frac{1}{2}$ electron operator and $\mathcal{U}(\bm{r})$ is a superlattice potential with lattice spacing $L$. A standard geometry, shown in Fig. \ref{f:chirpa}a and b, involves an antidot array which produces an electrostatic potential with maxima forming a triangular lattice and minima forming a honeycomb lattice 	
\cite{Sushkov2013,Tkachenko2015}. The electron density is depleted at the antidot sites and concentrated at the minima of the potential (Fig. \ref{f:chirpa}d). We assume the system is distanced from gates by $D\gg k_F^{-1}$. The precise form of the potential is subject to the particular design of the system, however these details are not conceptually important to our mechanism and we may treat a generic system via a simplified model involving three reciprocal lattice vectors $\bm{G}_1 = \frac{2\pi}{3 L} \left(3,\sqrt{3} \right)$, { $\bm{G}_2 = \frac{2\pi}{3 L} \left(0, -2\sqrt{3}\right)$, $\bm{G}_3 =  \bm{G}_1 + \bm{G}_2$,} and parametrized by an effective energy constant $W$,
\begin{align}
\mathcal{U}(\bm{r}) = 2W\sum_{i}{\cos(\bm{G}_i\cdot \bm{r})} 
\label{potential} \ .
\end{align}

As in graphene, the band structure features two band crossings at the valley momenta { $\bm K_1=\frac{4\pi}{3L}(1,0)$} and { $\bm K'_1=-\bm K_1$}, near which the single--electron dynamics is described by the Dirac Hamiltonian
\begin{align}
\label{dirac}
\mathcal{H}_0= \sum_{\bm{k}}{\psi^\dagger_{\bm{k}} v(\tau^z k_x\sigma^x+k_y\sigma^y) \psi_{\bm{k}}} \
\end{align}where $v$ is the effective velocity, $\psi_{\bm{k}}$ is an 8-component spinor possessing spin, valley ($\bm \tau$) and an additional pseudospin ($\bm \sigma$) degrees of freedom. The eigenstates of the pseudospin operator $\sigma^z$ correspond to electronic states with charge density residing primarily on either the $A$ or $B$ sublattices of the honeycomb structure surrounding the antidot sites.

The wavefunctions of (\ref{hamiltonian}) allow a direct computation of the matrix elements of the unscreened Coulomb interaction, in the basis of valley and pseudospin. The interactions near the Dirac points are
\begin{gather}
\label{bareint}
{\cal H}_{int} = \frac{1}{2} \sum_{\bm{k},\bm{p},\bm{q}}{ V_{\mu \nu; \rho \lambda}(\bm{q}) \left(\psi^\dagger_{\bm{k}+\bm{q}} \sigma^\mu \tau^\rho \psi_{\bm{k}}\right) \left(\psi^\dagger_{\bm{p}-\bm{q}} \sigma^\nu \tau^\lambda \psi_{\bm{p}} \right) } 
\end{gather}where the pseudospin and valley indices run over $\mu, \nu, \rho, \lambda \in \{0,x,y,z \}$ (with $\sigma^0$ and $\tau^0$ denoting the identity operator in pseudospin and valley space).

The interaction (\ref{bareint}) contains both the long range $1/q$ repulsion as well as additional short--ranged pseudospin and valley dependent repulsions $V_{\mu\nu;\rho\lambda}\propto 1/K$. We consider the situation where the chemical potential is tuned close to the Dirac points $k_F \ll K$, so the single--electron dynamics is well described by the Dirac theory. The pseudospin-dependent interactions at the Fermi surface are therefore suppressed compared to $1/q$ by a factor $k_F/K \ll 1$. 

The values of the interaction constants $V_{\mu \nu; \rho \lambda}$ are readily controlled by the ratio $W/E_K$ where $E_K = K^2/2m^*$. The system can be driven between two regimes: at low values of $W/E_K$, the Bloch wavefunctions correspond to nearly free electrons that reflect off the repulsive sites occurring at the maxima of $\mathcal{U}(\bm{r})$, and therefore occupy a significant portion of the unit cell. For larger values of $W/E_K$, the charge density becomes exponentially localized at the minima of $\mathcal{U}(\bm{r})$, which form a honeycomb lattice consisting of two well-defined interlocking sublattices (Fig. \ref{f:chirpa}d). We observe from numerics that this regime corresponds to $W/E_K > 1$. In order to understand the behaviour of the interaction constants in this regime, we may express Hamiltonian in the basis of Wannier orbitals $\varphi_{\alpha}(\bm{r})$, which we may restrict to the two orbitals nearest to the Fermi level,
\begin{widetext}
\begin{gather}
H = \sum_{\bm{R}_i, \alpha_i}{ T_{\alpha_1 \alpha_2}(\bm{R}_1-\bm{R}_2) c^\dagger_{\alpha_1,\bm{R}_1} c_{\alpha_2,\bm{R}_2} }
+ \frac{1}{2}\sum_{\alpha_i,\bm{R}_i}{
U_{\mu \nu}(\bm{R}_1, \bm{R}_2, \bm{R}_3, \bm{R}_4)\left( c^\dagger_{\alpha_1,\bm{R}_1} \sigma^\mu_{\alpha_1 \alpha_2} c_{\alpha_2,\bm{R}_2}\right) \left(c^\dagger_{\alpha_3\bm{R}_3}\sigma^\nu_{\alpha_3 \alpha_4} c_{\alpha_4,\bm{R}_4}\right)
}
\label{hamiltb}
\end{gather}
\end{widetext}
where $c^\dagger_{\alpha,\bm{R}} = \int{ \Psi^\dagger(\bm{r}) \varphi_\alpha(\bm{r}-\bm{R}) d\bm{r}}$, $\sigma^0 = 1$ and $\sigma^x, \sigma^y, \sigma^z$ are the Pauli matrices acting on the orbital doublet. We treat the creation operators $c^\dagger_{\alpha,\bm{R}}$ as two component-spinors in spin space, so summation over spins is implicit in the products $c^\dagger_{\alpha,\bm{R}} c_{\alpha',\bm{R}'}$.

For $W/E_K >1$, the orbitals become localized at one of the two minima of $\mathcal{U}(\bm{r})$ within a unit cell and we may consider $\alpha_i$ to index the two distinct sublattices indicated in red and blue in Fig. \ref{f:chirpa}d. Writing $c^\dagger_{\alpha,\bm{R}} \rightarrow (a^\dagger_{\bm{R}},b^\dagger_{\bm{R}})_\alpha$, the dominant single particle hopping processes in (\ref{hamiltb}) then occur when $\bm{R}_1$ and $\bm{R}_2$ are nearest neighbours (corresponding to terms of the form $a^\dagger_{\bm{R}_1} b_{\bm{R}_2}$, with other hoppings decaying exponentially with the distance between sites. At the same time, interaction terms in (\ref{hamiltb}) in  which $\alpha_1 \neq \alpha_2$, $\alpha_3 \neq \alpha_4$ (corresponding to terms of the form $a^\dagger_{\bm{R}_1} b_{\bm{R}_2} b^\dagger_{\bm{R}_3} a_{\bm{R}_4}$) are suppressed for the same reason. The remaining terms $\propto U_{00}, U_{zz}$ are diagonal in pseudospin space, and correspond to terms of the form $(a^\dagger_{\bm{R}} a_{\bm{R}} \pm b^\dagger_{\bm{R}} b_{\bm{R}})(a^\dagger_{\bm{R}'} a_{\bm{R}'} \pm b^\dagger_{\bm{R}'} b_{\bm{R}'})$. Both $U_{00}, U_{zz}$ are strongly enhanced when the orbitals become localized, { and as the superlattice potential becomes more confining, the short-ranged interactions with $\bm R = \bm R'$ become dominant.} The constants $V_{\mu \nu; \rho \lambda}$ describing interactions near the $K,K'$ points may be obtained by performing a Fourier transform with respect to $\bm{R}-\bm{R}'$.

Therefore, in the low-energy model (\ref{bareint}), the only surviving interactions in the regime $W/E_K > 1$ are diagonal in pseudospin space -- containing $\sigma^0$, $\sigma^z$ -- and can either contain $\tau^0, \tau^z$ (intravalley scattering) or $\tau^x, \tau^y$ (intervalley scattering). Both interactions can lead to superconductivity with different orders, but numerical results revealed that the intervalley mechanism yielded an instability for lower values of $T_c$ throughout all the range of parameters studied; we will exclude this from the subsequent analysis in the current work, since it is less relevant to experiment. The only interactions that we will therefore retain in our model are associated with intravalley, pseudospin-conserving scattering, ie $V_{00;00}(\bm q)$ and $V_{zz;00}(\bm q)$ in Eq. \eqref{bareint}. The pseudospin-independent interaction contains the long range Coulomb interaction, $V_{00;00}(\bm{q}) \rightarrow 2\pi e^2/\varepsilon_r |\bm{q}|$ for small $q$, while the pseudospin-dependent interaction $V_{zz;00}(\bm q)$ is short ranged (see Eq. \eqref{beta} of the Supplementary Material), { and is therefore an increasing function of $W/E_K$, shown explicitly in the Supplementary Material.}

We note furthermore that since we are interested in the regime $W/E_K > 1$, in which the Wannier orbitals occupy a small region of the unit cell, our analysis is not applicable to graphene in which the atomic radius is comparable to the lattice spacing $L = 2.54$\r{A}. The ability to separate the length scales associated with the size of the orbital wavefunctions and the lattice spacing in AG allows access to dramatically different interaction effects.

\section{Pairing Mechanism}
\label{mech}
There are two key aspects of the mechanism for superconductivity. The first arises due to the novel screening properties of the pseudospin-dependent Hubbard interactions we introduce in Eq. \eqref{bareint}, and the second arises due to the topological properties of the Dirac wavefunction which results in an effective attraction. 

We firstly incorporate screening in the Random Phase Approximation (RPA) \cite{GonzalezN,SonN,Hwang2007,Gangadharaiah2008,Kotov2008,KotovN,Hofmann2014}, which can be justified here by the fact that the Dirac cones have fourfold ($N=4$) spin and valley degeneracy; corrections to the RPA are relatively suppressed by the factor $1/N$ \cite{GonzalezN,SonN,KotovN}.  We comment that unlike ordinary metals in which the RPA is an expansion in the Wigner--Seitz radius $r_s$, which is typically not a small parameter, in Dirac systems the RPA is an expansion in $1/N$, making the RPA well-controlled \cite{DasSarma2011}. While previous studies have applied the RPA to describe screening effects in graphene, in our current analysis we also incorporate the pseudospin-dependent interaction $V_{zz;00}$ in addition to the usually considered long range $\propto 1/q$ interaction. The general form of the RPA equations for the interaction structure (\ref{bareint}) is \begin{gather}
\widetilde{V}_{\mu \nu; \rho \lambda}(\omega, \bm{q}) = \nonumber \\
 V_{\mu \nu; \rho \lambda}(\bm{q}) + V_{\mu \alpha; \rho \beta}(\bm{q}) \Pi^{\alpha \gamma; \beta \delta} (\omega,\bm{q}) \widetilde{V}_{\gamma \nu; \delta \lambda} (\omega, \bm{q})
\end{gather} where the polarization operators $\Pi^{\alpha \gamma; \beta \delta}$ are given by 
\begin{gather}
\Pi^{\alpha \gamma; \beta \delta}(\omega, \bm{q}) = \nonumber \\
-i \text{Tr} \int{
\sigma^\alpha \tau^\beta G(E+\omega, \bm{k}+\bm{q}) \sigma^\gamma \tau^\delta G(E, \bm{k}) \frac{ dE d^2\bm{k}}{(2\pi)^3}
} \ \ , \nonumber \\
G(E, \bm{k}) = \frac{1}{ E+\mu - v ( \tau^z k_x \sigma^x + k_y \sigma^y) +i 0 \text{sgn}(E)}
\label{Pi}
\end{gather} 
with $G(E, \bm{k})$ being the single particle Green's function. Note that the intervalley and intravalley interactions are always screened independently due to the structure of the trace in (\ref{Pi}).  Solution of the RPA equations and evaluation of (\ref{Pi}) show that $\Pi^{00;00}$ in our case is identical to the  polarization operator of graphene \cite{Hwang2007}, which leads to the screening of the long range Coulomb repulsion. By contrast, the expression for $\Pi^{zz;00}$ contains a Pauli operator $\sigma^z$ which anticommutes with the single-particle Hamiltonian $v(\tau^z k_x \sigma^x + k_y \sigma^y)$ in the denominator of the Green's function, leading to a different screening phenomenon. In particular, in the range $0 < |\bm{q}| < 2k_F$ the static polarization operators $\Pi^{00;00}$, $\Pi^{zz;00}$ are constant and real, but opposite in sign, leading to the reduction and enhancement respectively of $V_{00;00}$, $V_{zz;00}$:
 \begin{gather}
\Pi^{00;00}(\omega = 0, \bm{q}) = -\frac{2 k_F}{\pi v} \nonumber \\
\Rightarrow \widetilde{V}_{00;00}(\omega=0,\bm{q}) =  \frac{ V_{00;00}(\bm{q}) }{1 + \frac{2 k_F}{\pi v} V_{00;00}(\bm{q})} \ \ ,
\end{gather} 
while
 \begin{gather}
\Pi^{zz;00}(\omega = 0, \bm{q}) = + \frac{2k_F}{\pi v} \nonumber \\   \Rightarrow\widetilde{V}_{zz;00} (\omega = 0,\bm{q}) = \frac{ V_{zz;00}(\bm{q})}{1 - \frac{2k_F}{\pi v} V_{zz;00} (\bm{q})} \ \ .
\end{gather}
The complete expression for $\Pi^{zz;00}(\omega,\bm{q})$ at finite frequency is given in the Supplementary Material.

\begin{figure}[t]
 \hspace{0.2 cm}
\raisebox{0.0\height}{\includegraphics[width=37mm]{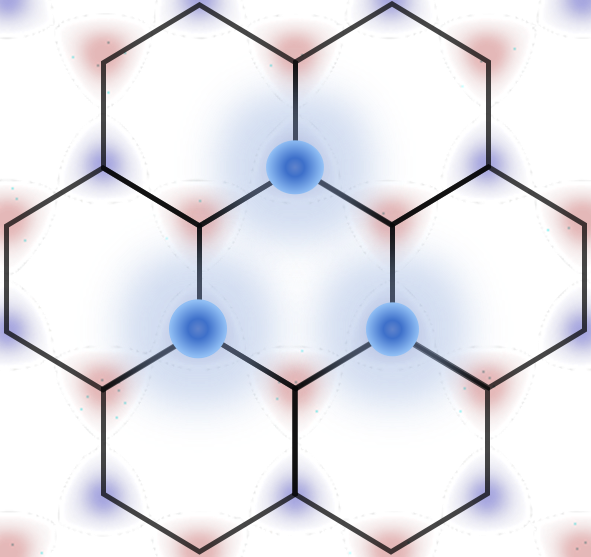}}\hspace{0.7 cm}
\raisebox{0.0\height}{\includegraphics[width=37mm]{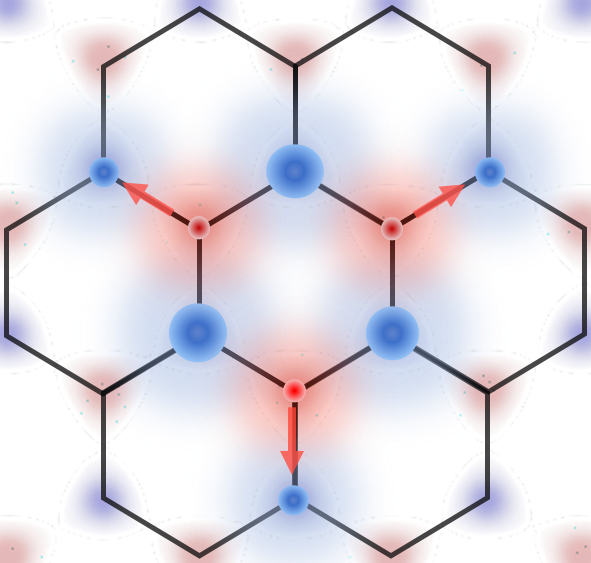}\hspace{0.0 cm}}
\vspace{0.3 cm}
\begin{picture}(0,0) 
\put(-250,103){\textbf{(a)}} 
\put(-125,103){\textbf{(b)}}

\put(-188,46){$ \mathlarger{\mathlarger{\mathlarger{ \sigma^z }}}$ }

\put(-60,46){$ \mathlarger{\mathlarger{\mathlarger{ \sigma^z }}}$ }

\put(-25,46){$ \mathlarger{\mathlarger{\mathlarger{ \sigma^z }}}$}
\put(-94,46){$ \mathlarger{\mathlarger{\mathlarger{ \sigma^z }}}$ }

\put(-42,75){$ \mathlarger{\mathlarger{\mathlarger{ \sigma^z }}}$ }
\put(-77,75){$ \mathlarger{\mathlarger{\mathlarger{ \sigma^z }}}$}

\put(-42,17){$ \mathlarger{\mathlarger{\mathlarger{ \sigma^z }}}$ }
\put(-77,17){$ \mathlarger{\mathlarger{\mathlarger{ \sigma^z }}}$ }
\end{picture}
\vspace{-0.3 cm}
\caption{{\textbf{Pseudospin antiscreening} Electrons (blue) are deposited on the $A$ sublattice in \textbf{(a)} locally increasing the pseudospin $\langle \sigma^z\rangle = n_A-n_B$. Surrounding electrons are pushed off the neighbouring $B$ sublattice in \textbf{(b)} leaving holes (red) behind, increasing the pseudospin in the neighbouring unit cells -- generating a ``ferromagnetic'' pseudospin polarization. The effect is to enhance the initial $\langle \sigma^z \rangle$, which corresponds to increasing the effective $V_{zz;00}$ coupling. This is contrasted with the overall charge density $n_A+n_B$, which is screened.
}  \label{f:antiscreening}} 
\end{figure}
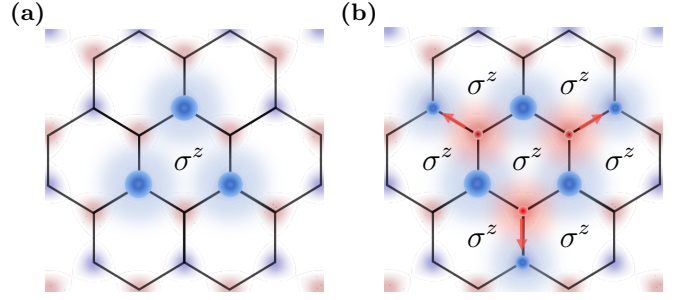

The function $\Pi^{zz;00} > 0$ implies $V_{zz;00}$ is antiscreened, i.e. the screened interaction is enhanced relative to its the bare value. There is a simple physical picture for why this enhancement occurs. Conventional charge screening occurs because upon placing an electron in a metal, neighboring electrons are repelled creating a cloud of holes around it, so that at long distances an observer sees an effective charge which is smaller than the bare  charge due to shielding. By contrast, consider the response of the system to a local imbalance of charge between the two sublattices (ie a local increase in the pseudospin $\sigma^z$). The  imbalance pushes charge off the neighboring sites, which in turn causes a charge imbalance in the neighboring unit cell, as illustrated in Fig. \ref{f:antiscreening}. Hence, the total sublattice imbalance $\sigma^z$ an observer sees at long distances has been \textit{increased}. This means that pseudospin fluctuations are antiscreened, and grow stronger due to many body effects. Formally, this effect can also be derived from the anticommutation relation between the  operator $\sigma^z$ and the kinetic term $\propto \tau^z k_x\sigma^x + k_y\sigma^y$. Physically, $\Pi^{zz;00}$ measures the response of the system to a local perturbation of $\langle\sigma^z\rangle$, the relative electron density on sublattices $A$ and $B$. The result $\Pi^{zz;00} > 0$ therefore indicates that a such a perturbation induces a ``ferromagnetic" pseudospin polarization of the neighboring environment, as illustrated in Fig. \ref{f:antiscreening}.

To investigate the possibility of superconductivity, it is necessary to calculate the scattering amplitude in the  Cooper channel, using the Dirac wavefunctions for particles on the Fermi surface,
 \begin{align}
 |\bm k,\tau\rangle = \frac{1}{\sqrt{2}}e^{i\bm k \cdot \bm r} (|a\rangle + \tau e^{i \tau \theta_{\bm{k}}} |b\rangle ) \ \ ,
\label{Diracwf}
 \end{align} with $|a \rangle,|b\rangle$ being the $\sigma^z$ eigenstates, which are localized on the $A$ and $B$ sites respectively. The matrix elements of the interactions between the outgoing and incoming states $\bm k, -\bm k$ and $\bm p, -\bm p$ within the same valley result in the scattering amplitude 
\begin{align}
\label{intra}
\Gamma_{\tau\tau}(&\bm{p},\bm{k})=\frac{1}{4}\widetilde{V}_{00;00}(\omega,\bm q)\left({1+e^{i\tau\theta}}\right)^2 \nonumber \\
&+ \frac{1}{4}\widetilde{V}_{zz;00}(\omega,\bm q)\left({1-e^{i\tau\theta}}\right)^2
\end{align}where $\bm{q} = \bm{k}-\bm{p}$, $\omega = v( |\bm{k}| - |\bm{p}|)$ and $\theta=\theta_{\bm p}-\theta_{\bm k}$ is the scattering angle. The dominant contributions to the BCS equations arise from scattering processes near the Fermi level, for which the $\omega$ and $\bm{q}$ dependence of $\widetilde{V}_{00;00}$ and $\widetilde{V}_{zz;00}$ is weak, and we may replace them with constants. The momentum dependence of the scattering vertex then appears  entirely due to the scattering angle in terms containing $e^{i \tau \theta}$ in (\ref{intra}), which originate from the Berry phase of the Dirac wavefunctions. Decomposing the interaction into partial waves we find the amplitudes, 
\begin{align}
\label{partial}
\Gamma_{\tau \tau}^{\ell} = \int{ \Gamma_{\tau \tau}(\bm{p},\bm{k}) e^{-i \ell \theta} \frac{d\theta}{2\pi}} =\ \ \ \ \ \ \ \ \ \ \ &
\nonumber \\
\begin{cases}
\ \frac{1}{4}\left(\widetilde{V}_{00;00}+ \widetilde{V}_{zz;00}\right) \  ,  & l = 0, 2\tau, \\
\ \frac{1}{2}\left(\widetilde{V}_{00;00}- \widetilde{V}_{zz;00}\right) \  ,  & l = \tau, \\
\ 0 \ \ , &l \neq 0, \tau, 2\tau.   
\end{cases}
\end{align}

From equations \eqref{intra} and \eqref{partial}, we see that the term associated to $\widetilde{V}_{zz;00}$ has an attractive $\ell = \pm 1$ partial wave amplitude due to the negative prefactor of $e^{\pm i \theta}$. The interaction promotes $p+ip$ pairing in the valley $\tau=1$ and $p-ip$ pairing in the other valley $\tau=-1$, which we denote as $p+i\tau p$ pairing. When $\widetilde{V}_{zz;00} > \widetilde{V}_{00;00}$, this attraction exceeds the residual repulsive interactions, resulting in $\Gamma^{\ell=\tau}_{\tau\tau}<0$ and hence superconductivity.

The attraction results from the interplay between the kinetic term in the Hamiltonian and interaction-driven fluctuations of the relative charge between the sublattices $\langle \sigma^z \rangle = n_A - n_B$, where $n_A$ and $n_B$ are the densities on each sublattice. The Dirac wavefunctions are equal superpositions of $\sigma^z$ eigenstates with a relative phase which winds in momentum space; the matrix elements associated with both $\widetilde{V}_{00;00}$ and $\widetilde{V}_{zz;00}$ inherit this phase winding when projected onto the Fermi surface $(n_A \pm n_B)^2 \rightarrow \frac{1}{4} ( 1 \pm e^{-i \tau \theta})^2 n_+^2$ (with $n_+$ being the density in the upper band). Hence, while the interactions are proportional to $(n_A \pm n_B)^2$ which is always positive, the phase winding of the Dirac wavefunction causes the interaction to separate into different angular harmonics with opposite sign.

\begin{figure}[t]

\includegraphics[]{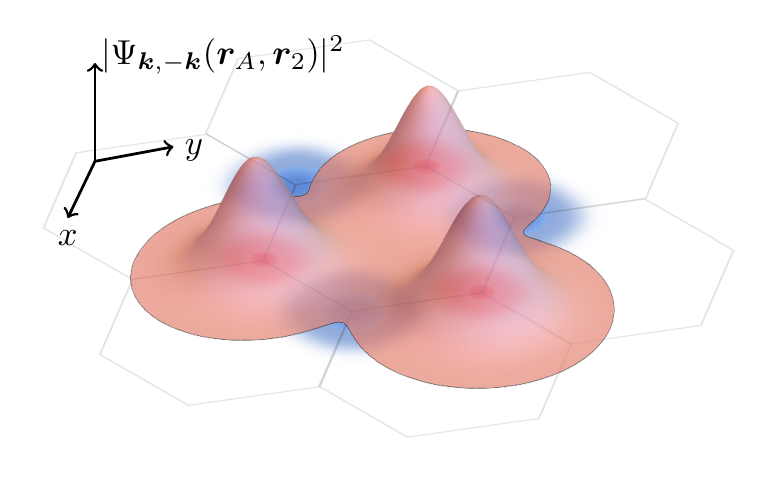}

\vspace{-0.1cm}

\caption{{\textbf{Interference between Dirac particles} { The relative phase between Dirac particles with momenta $\bm k$ and $-\bm k$ means that a pair wavefunction made out of these states acquires nodes on the lattice sites. This minimizes the repulsive energy cost of the pseudospin (sublattice polarization, c.f. Fig. \ref{f:antiscreening}) fluctuations $\sim V_{zz;00}$, by causing the electrons to avoid being on the same sublattice, and therefore reducing the total pseudospin $\langle \sigma^z \rangle =n_A-n_B$. Antiscreening enhances this energy gain, which eventually gives rise to superconductivity. The nodal structure is illustrated by a schematic plot of the two--electron probability density $|\Psi_{\bm k, -\bm k}(\bm r_1, \bm r_2)|^2$, where the coordinate of one electron is fixed to $\bm r_1 = \bm r_A$ on sublattice $A$ (blue sites). The probability density for $\bm{r}_2$ then has maxima on the $B$ sites (red sites) and nodes on the $A$ sites.}
}  \label{f:nodes}} 
\end{figure}

The pairing mechanism may be explained more intuitively as follows. While the unscreened Coulomb interaction is always repulsive, it is reduced when electrons avoid being on the same sublattice. For a Dirac particle, the wavefunction acquires a $\pi$ Berry phase when $\bm k \rightarrow -\bm k$. This implies that a pair wavefunction made out of two particles, with momenta $\bm k$ and $-\bm k$, acquires nodes due to destructive interference between the two components (Fig. \ref{f:nodes}); if one electron is located on the $A$ sublattice, its pair is forced to live on the $B$ sublattice. Since the interaction associated to $\widetilde{V}_{zz;00}$ is $\propto (n_A-n_B)^2$, forming a state out of pairs with $\bm k$ and $-\bm k$ causes the electrons to minimize the energy cost of $n_A\neq n_B$, by avoiding occupying the same sublattice.  The energetic advantage from forming such a state with these nodes $\sim \widetilde{V}_{zz;00}$, and when antiscreening causes this energy to exceed the repulsion from the average charge density $\sim \widetilde{V}_{00;00}$, the system minimizes energy by forming Cooper pairs.

The combination of the negative $\ell = \pm 1$ ampltude in the $\widetilde{V}_{zz;00}$ interaction due the interference associated with Berry phase, and antiscreening, is a new mechanism for superconductivity due to repulsive interactions, which has not been previously explored in other studies of honeycomb lattice models. Furthermore we note that, unlike the most commonly studied routes to superconductivity via repulsive interactions, pairing does not rely on nesting \cite{Nest}, spin fluctuations \cite{Honerkamp2008, Ma2011,Wu2013,Vladimirov2019}, van Hove singularities \cite{McChesney2010,Kiesel2012,Nandkishore2012}, or the Kohn--Luttinger mechanism -- which relies on singularities in the interaction due to backscattering, ie scattering processes with $q=2k_F$ \cite{Nest,Kagan2015}. Our mechanism also exists at weak coupling, and the `pairing glue' is the fluctuations of the pseudospin density $\langle \psi^\dagger \sigma^z\psi\rangle$.

In Figure \ref{f:results}a we plot the $\ell=\tau$ partial wave amplitude $\Gamma^{\ell = 1}_{\tau\tau}(p,k_F) $. The low--frequency attractive part of the intravalley interaction is due to the antiscreened $\widetilde{V}_{zz;00}$ term, while at high--frequencies, both the screened and antiscreened interactions are repulsive. The same step--like frequency dependence appears in phonon--mediated BCS pairing, for which the interaction consists of a screened repulsive part (the so--called Anderson--Morel pseudopotential) and an overscreened attractive part beneath the Debye frequency due to phonons \cite{Morel1962,RetSC}. We therefore solve the gap equation using standard methods from BCS theory (for details see the Supplementary Material).

 \section{Superconducting Gap and Critical Temperature}
\label{gap}

\begin{figure*}[t]
\hspace{-0.0cm}
\raisebox{-0.018\height}{\includegraphics[width=57.8mm]{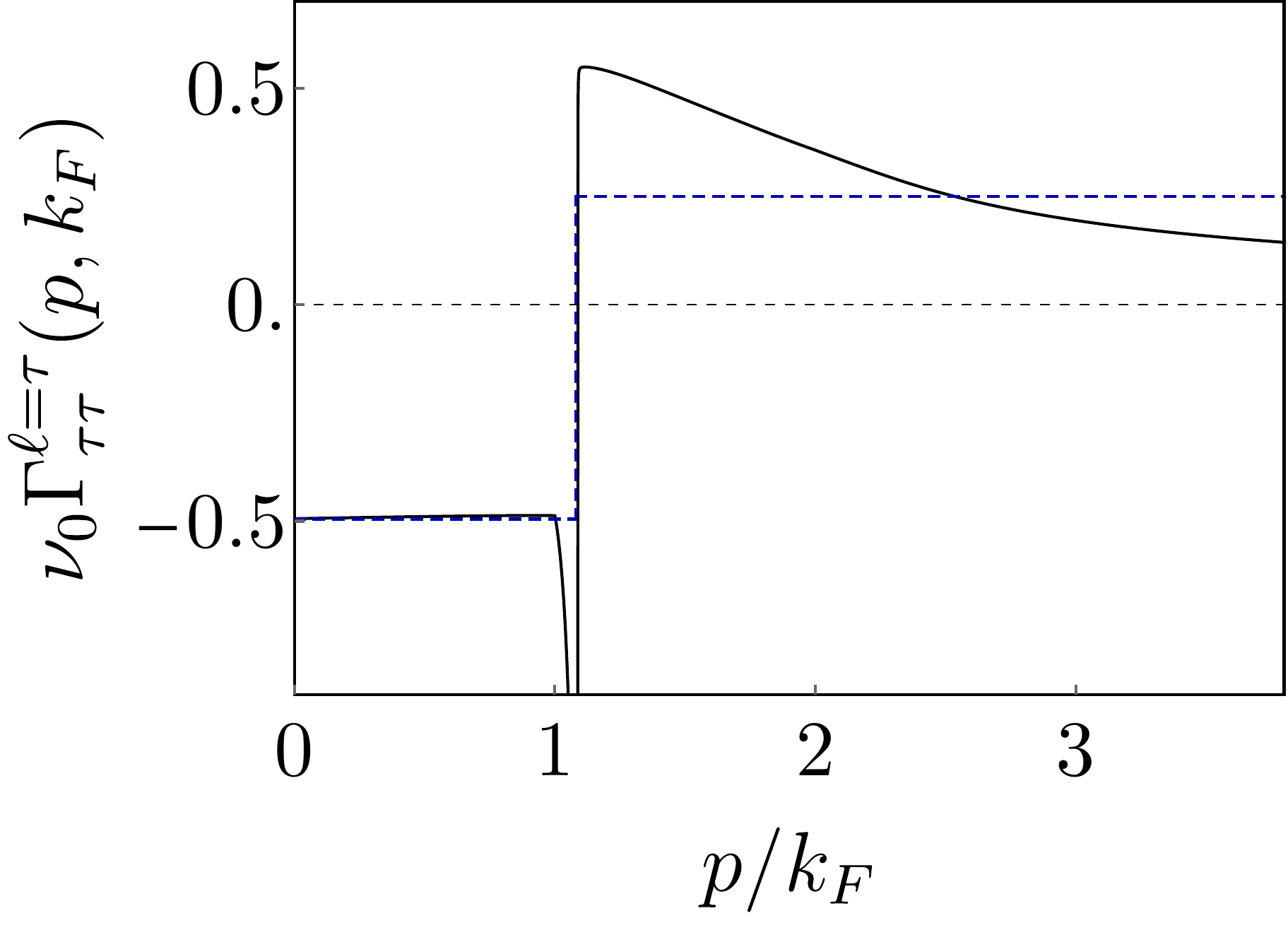}}
\raisebox{-0.00\height}{\includegraphics[width=54mm]{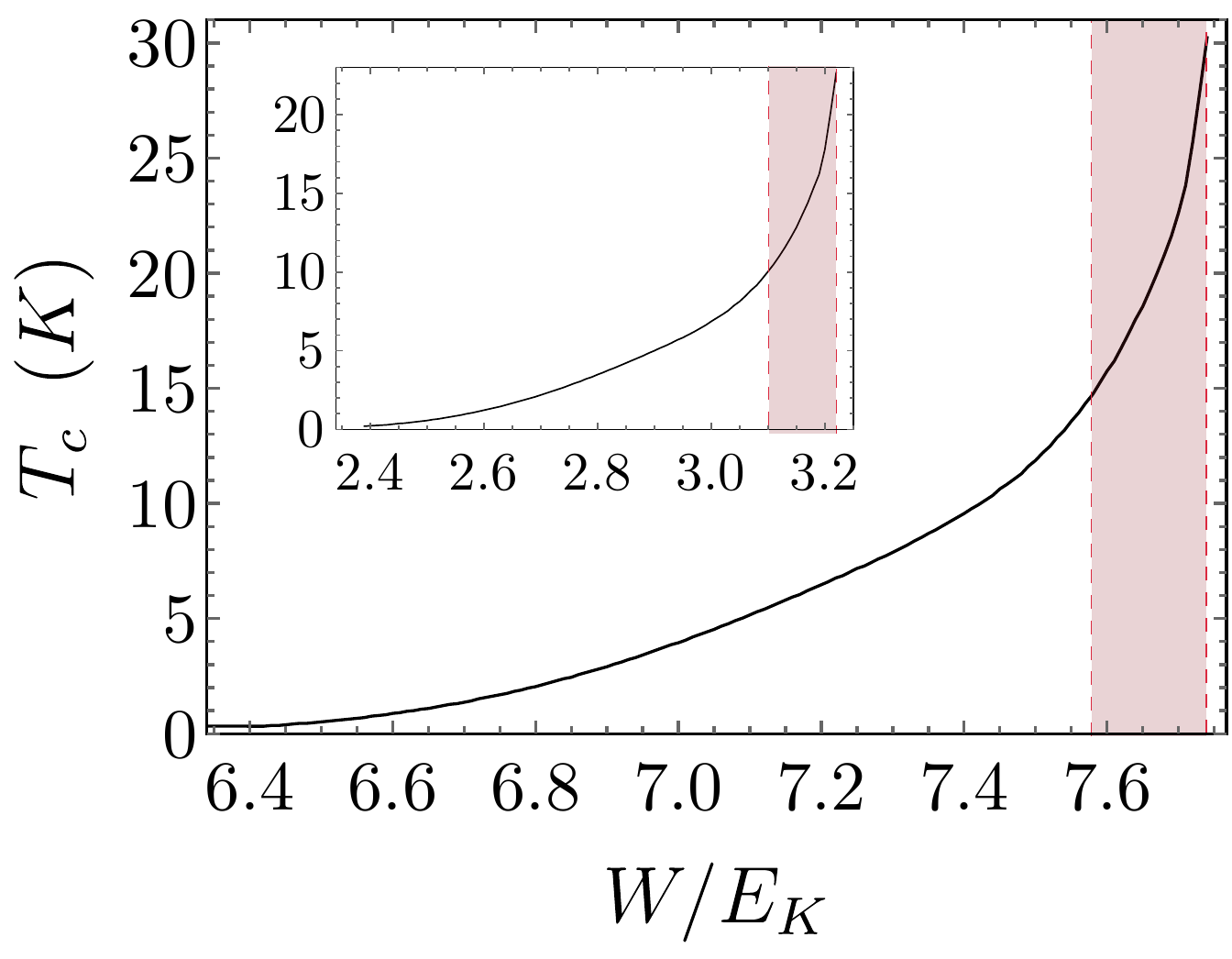}}
\raisebox{0.012\height}{\includegraphics[width=62.5mm]{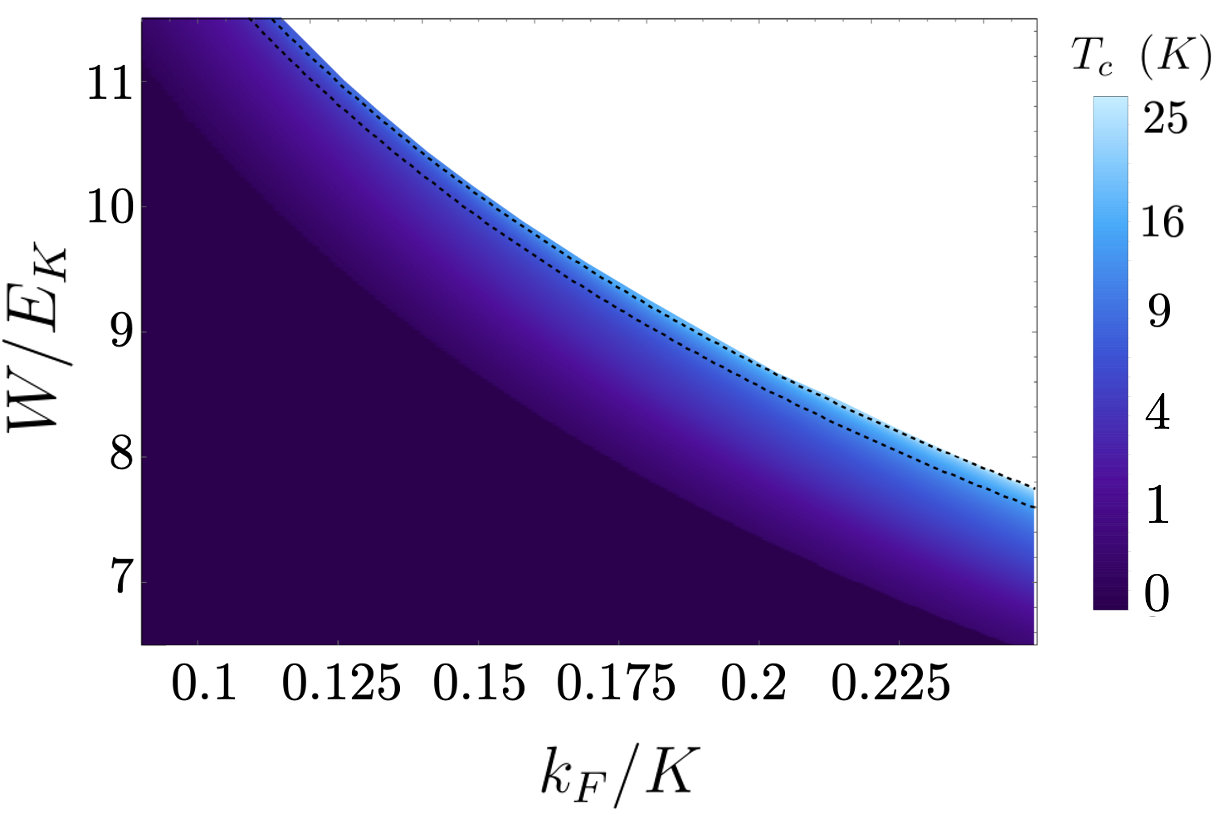}}
\begin{picture}(0,0) 
\put(-487,120){\textbf{(a)}} 
\put(-338,120){\textbf{(b)}} 
\put(-178,120){\textbf{(c)}} 
\end{picture}
\caption{ {{\textbf{(a)} The $\ell=\tau$ partial wave component of the Cooper channel scattering amplitude (\ref{intra}), $\nu_0\Gamma^{\ell=\tau}_{\tau \tau}(p,k_F)$ for parameters corresponding to a $10$ nm InAs quantum well with $W/E_K = 7.3$ and $k_F/K=0.25$. The solid line indicates the full momentum dependence in Eq. (\ref{intra}), while the blue dashed line is the step function approximation Eq. (\ref{retarded}).  \textbf{(b)} Critical temperature as a function of $W$ for InAs and GaAs (inset) quantum wells at doping $k_F/K=0.25$. The red shaded region indicates that the ratio of gap to Fermi energy $0.1<\Delta/E_F<0.2$. \textbf{(c)} Heat plot of critical temperature as a function of Fermi momentum $k_F/K$ and potential strength $W/E_K$. The lower (upper) dashed line marks $\Delta/E_F=0.1 (0.2)$, while the unshaded region $\Delta/E_F>0.2$ is strongly coupled and contains competing instabilities.}  \label{f:results} } }
\end{figure*}
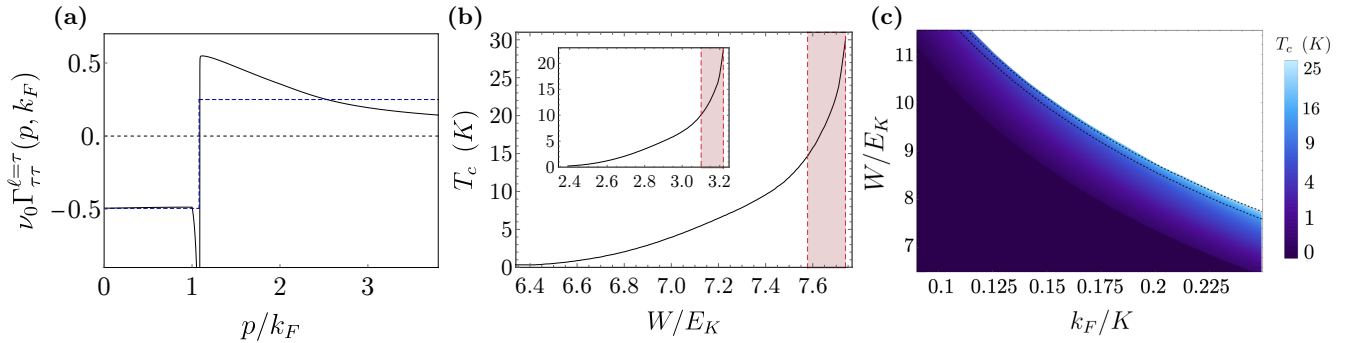

As shown in Figure \ref{f:results}a, the (dimensionless) intravalley scattering amplitude can be well approximated by a simplified form consisting of step functions in $k,p$ with three positive parameters ($g_{1,2,3}$),
\begin{align}
\label{retarded}
\nu_0\Gamma^{\ell = \tau}_{\tau \tau}(p,k) = \begin{cases}
 g_2 \Theta(p - \Omega) -g_1 \Theta(\Omega - p) \  , \  k < \Omega \\
g_2 \Theta(\Omega - p) - g_3 \Theta(p - \Omega) \  , \ k >\Omega 
\end{cases} 
\end{align} where $\nu_0$ is the single--particle density of states at the Fermi energy $E_F$, measured relative to the Dirac point. The parameter $\Omega$ is the frequency at which the scattering amplitude changes sign (see Fig \ref{f:results}a), and in the regime we consider is comparable to the Fermi energy. The quantities $g_i$ are obtained by averaging the dimensionless scattering amplitude $\nu_0\Gamma$ for frequencies above and below $\Omega$. An attractive interaction between electrons within the same valley in the $\ell = \tau$ channel implies a spin triplet $p+i \tau p$ superconducting gap,
\begin{gather}
\Delta_{s,s';\tau}(\bm{k}) =\sum_{\bm{p}}{
	\Gamma^{\ell=\tau}_{\tau \tau}(p,k) e^{i\tau ( \theta_{\bm{p}}  - \theta_{\bm{k}})} \langle \psi_{-\bm{p},s,\tau} \psi_{\bm{p},s',\tau}\rangle 
} \nonumber \\
= \Delta^{\ell = \tau}(k) k^{-1}(k_x- i\tau k_y) (i\bm{d}\cdot \bm{s} s_y)_{s s'}
\label{gapsymmetry}
\end{gather}
where $\bm{d}$ is a real three--dimensional vector associated with the spin triplet ordering, $\bm{s}$ are the Pauli matrices acting on spin, and $\Delta^{\ell = 1}$ satisfies the BCS gap equation,

\begin{gather}
\Delta^{\ell=\tau}_\tau(k)  = \nonumber \\
\int{
	\frac{-\Gamma^{\ell=\tau}_{\tau \tau}(p,k) \Delta^{\ell = 1}(p)}{ 2\sqrt{ \epsilon_{p}^2 + |\Delta^{\ell=1}(p)|^2}} \tanh \frac{ \sqrt{ \epsilon_p^2 + |\Delta^{\ell=1}(p)|^2}}{ 2T} \frac{d^2p}{(2\pi)^2}
} 
\label{bcs}
\end{gather}the solution of which is given in the Supplementary Material. The gap is a superposition of $p+ip$ pairing in one valley, and $p-ip$ in the other. We also note that intravalley pairing has the novelty of a gap with nonzero quasimomentum $K$, which has the form $\Delta(\bm{r}) \propto \cos (2\bm{K}_1\cdot \bm{r} + \phi)$, i.e. spatially modulated with a periodicity of three unit cells with $\phi$ being a constant phase, realizing an example of the unusual Fulde--Ferrell--Larkin--Ovchinnikov (FFLO) phase (alternatively a `pair density wave') \cite{
Fulde1964,Larkin1965,Roy2010,Tsuchiya2016}.

In Figure \ref{f:results}b, we fix the { electronic density} at $k_F/K=0.25$, and plot $T_c$ against potential strength $W/E_K$ where $E_K = K^2/2m^*$ for both InAs, and GaAs (inset) based AG. Taking $L=10$ nm, for InAs with $\varepsilon_r=14.6$ and $m^*=0.0229$ \cite{Winkler2003} we find $T_c\approx 20$ K. For GaAs, with $\varepsilon_r=12.4$ and $m^*=0.067$, for $L=10$ nm we find $T_c\approx10$ K. The shaded regions on the two curves indicate entry into the strong coupling regime, determined by $0.1<\Delta/E_F<0.2$. We note that $T_c$ for GaAs is lower than in InAs, due the larger effective mass and hence lower $E_K$. The ratio between the gap and Fermi energy $\Delta/E_F$ is commonly used as an indicator of the strength of the pairing \cite{Cao2018}. The value $\Delta/E_F = 0.1$ is comparable to many known high temperature superconductors, illustrating that the values of $T_c$ possible in AG are large in comparison to the low carrier density.

Unlike many existing theories of high temperature superconductors, weak coupling BCS theory can be reliably applied to our microscopic theory of pairing. We note that the fine structure constant in our theory is $\alpha = e^2/(\varepsilon_r v) \approx 0.4$ for the parameters corresponding to the results shown for InAs in Fig. \ref{f:results}b and $\alpha \approx 0.7$ in GaAs (in the inset). For doping significantly away from the Dirac point, the strong screening of interactions due to the high degeneracy ($N=4$) of the Fermi surfaces means that $\alpha$ is less relevant as a measure of the interaction strength. Nevertheless we comment that, despite  a high value of $\alpha$ in graphene, it has been shown that the RPA is well controlled in graphene even at charge neutrality where the density of states is vanishing \cite{Hofmann2014}. In our current work, the relevant expansion parameter of perturbation theory is $(\log (2E_F/\Delta))^{-1}$ since perturbative corrections to the effective vertex are dominated by the logarithmic terms arising in the Cooper channel. 

In Figure \ref{f:results}c, we show $T_c$ as a function of density and potential strength using parameters for InAs, setting $L=10$ nm. For any doping within the range shown, there is a corresponding range of potential strengths for which superconductivity exists. This is an important result, demonstrating that superconductivity is a not a result of fine tuning. { Beneath a minimum $W/E_K$, the bare value of $V_{zz;00}$ becomes too small to yield superconductivity, despite antiscreening. } There is a minimum doping $k_F/K$ required to enhance the screening effects which cause $\Gamma^{\ell=\tau}_{\tau\tau}<0$, and we interrupt the phase diagram before antiscreening causes the couplings to grow large. 

{ For the unshaded regions in Fig. \ref{f:results}c, $\Delta/E_F > 0.2$.} The  dashed lines in the plots mark the region $0.1<\Delta/E_F<0.2$, at which point the interactions grow large, and perturbative corrections become uncontrolled. Nevertheless in this region, it is reasonable to expect superconductivity to persist and that $T_c$ will continue to increase, but our theory loses reliability.  In this regime, competing Stoner-type instabilities including ferromagnetism and density wave ordering begin to emerge. Despite the lack of theoretical predictions for this region of the phase diagram, we expect the system to exhibit interesting physics to be explored experimentally. Importantly, the Stoner--type instabilities do not appear at weak coupling because the Fermi surface is not nested, which is required for these instabilities to compete with superconductivity  in this regime \cite{Nest}. 

We note that the superconducting portion of the phase diagram requires a large ratio $W/E_K \gtrsim 6 \ (2.4)$ in InAs (GaAs), which implies $W \gtrsim$ 2 eV (0.25 eV). The large potential strength is a result of the simplified model \eqref{potential} with only a single energy parameter $W$, chosen for a simple conceptual illustration of the theory. In a situation where the superlattice is engineered to effectively increase the antidot size relative to the lattice spacing, additional cosine terms must be added in \eqref{potential} which preserve the lattice symmetry, and similar values of $V_{zz;00}$ may be obtained with a significantly shallower potential variation, as we illustrate in the Supplementary Material. 

\section{Discussion}
\label{disc}

 Microscopic calculations of $T_c$ typically cannot be taken as accurate experimental predictions, since $T_c$ depends exponentially on the electron-electron interaction and so modest errors in calculating the electron--electron scattering amplitude have an exponential effect on $T_c$. We nonetheless claim the above calculations are experimentally meaningful. The reason for this is the antiscreening of the $\widetilde{V}_{zz;00}$ coupling. As pointed out earlier, as one varies the chemical potential, lattice spacing, and potential strength -- as one is free to do in an artificial superlattice -- there is a region of parameter space where antiscreening causes $\widetilde{V}_{zz;00}$ to grow sufficiently large. Hence, if our approximations overestimate $\widetilde{V}_{zz;00}$, that error can be compensated for by experimentally varying the physical parameters so that $\widetilde{V}_{zz;00}$ is sufficiently enhanced by antiscreening for superconductivity to arise. Our results should then be interpreted as indicating the range of $T_c$ possible within a range of experimental parameters.

In order to understand the experimental feasibility of our proposal, it is important to ask whether our mechanism survives the disorder expected to be present in a nanofabricated device: impurity scattering, as well as superlattice disorder, i. e. shape, size and position variations of the antidots. In the case of both InAs and GaAs with $L=10$ nm, $\Delta/E_F =0.1$ and $k_F/K=0.25$ implies a superconducting coherence length $\xi = 30$ nm. On the other hand, the mean free path $l$ for these materials can be order several microns, substantially longer than the coherence length and the characteristic lattice spacing $l\gg \xi, L$, indicating that the effects of impurity scattering are weak and the superconducting state is described by the clean limit \cite{Mineev}. It has been shown that antidot size variation is the dominant long--wavelength superlattice disorder, and generates variations in the Fermi energy across the sample \cite{Tkachenko2015}. Even though the gap function \eqref{gapsymmetry} is time reversal symmetric, pairing does not occur between time--reversed states and Anderson's theorem \cite{Abrikosov1958,Anderson1959} does not apply, so we expect that $T_c$ is reduced by regions of the sample deviating from optimal doping \cite{Finkelstein1994}.  We finally note that it is possible for disorder to promote another superconducting state over $p+ i\tau p$ intravalley pairing -- Anderson's theorem would apply to $s$-wave intervalley pairing, for which we found lower $T_c$ in the absence of disorder. Given that the intervalley calculation yielded lower $T_c$, we will present the details in a future work.

Promisingly, very recent experimental work has reported the creation of low disorder AG, as indicated by clear signatures of the Dirac dispersion \cite{AG,AG2}. These low disorder realizations of AG possess superlattice spacings $L\sim 70$ nm; the main experimental challenge in realizing our predicted superconducting state with $T_c\approx 20$ K is achieving smaller values of $L\approx 10$ nm while maintaining low superlattice disorder. While we have plotted results for this ambitious scenario, superconductivity still exists for larger lattice spacings: for superlattices with $L=50$ nm our theory predicts up to $T_c\approx 1.4$ K (0.3 K) for InAs (GaAs) at the edge of the weak coupling regime $\Delta(T=0)=0.1E_F$.

Our theory naturally predicts an unconventional gap function (\ref{gapsymmetry}), corresponding to a $p+i\tau p$ FFLO state possessing time-reversal symmetry. Due to the valley and spin degeneracy, our state can be thought of as four copies of the Read-Green model of a 2D topological superconductor \cite{ReadGreen2000}. As pairing occurs within both the $K$ and $K'$ valleys, we expect two time-reversed pairs of Majorana edge modes, which  generally become hybridized. Such a state possesses a trivial topological invariant according to the bulk-boundary correspondence for topological superconductors, but may be shown to realize a higher-order topological phase with  protected zero-energy Majorana corner modes \cite{Peng2017,Langbehn2017}, a property which has major applications to topological quantum computation \cite{Zhang2020}. This point will be discussed further in a separate study.

While the superconducting phase discussed is among several explored previously in honeycomb lattices \cite{Uchoa2007,Zhou2013,Roy2010,Tsuchiya2016,Kunst2015}, our analysis demonstrates a new microscopic mechanism that gives rise to such a state: a pairing interaction mediated by antiscreened pseudospin fluctuations $\langle \sigma^z\rangle$, which exists generically as a feature of interacting Dirac systems. Previous field theoretic studies of honeycomb systems have explored the interaction $V_{00;00}$ and calculated the charge screening due to $\Pi^{00;00}$, but not the pseudospin interaction $V_{zz;00}$ -- which becomes essential in the regime where the radius of the atomic sites is small compared to the unit cell -- or antiscreening due to $\Pi^{zz;00}$.

While the mechanism is unlikely to be relevant to graphene, due to its relatively delocalized atomic orbitals and hence small value of $V_{zz;00}$, this mechanism might be applicable to other artificial lattices \cite{Kalesaki2014,Delerue2015,Boneschanscher2014,Gomes2012}, including recently discussed Moir\'{e} superlattices in twisted layered systems \cite{Yoo2019}, which would also have the advantage of much lower superlattice disorder. We stress, however, that  application of our theory would not require experimental fine tuning to the `magic angle'; our mechanism can exist in the weak coupling regime, as in phonon--mediated superconductivity, and does not require flat bands.

The tunability of AG presents the opportunity to test simple variations of this theory, in much the same way that cold atomic gases have allowed experimentalists a platform to implement a large host of toy models. Alternative lattice geometries can be imposed on the 2DEG, and future studies may also wish to investigate the role of higher bands beyond the first two Dirac points. An alternative avenue to chiral superconductivity in AG is to increase the density to a van Hove singularity, causing $d$-wave superconductivity alongside competing magnetic order, a scenario first proposed in the context of graphene \cite{McChesney2010,Kiesel2012,Nandkishore2012}. A crucial distinction in our mechanism is its validity over a large range of densities, and the absence of nesting and competing instabilities in the weak coupling regime. The ability to tune AG to the strong coupling regime, outside the parameter range for which we are confident superconductivity dominates, allows access to a significantly richer phase diagram in which density wave and magnetic order compete with chiral superconductivity -- an interesting scenario reminiscent of cuprates and twisted bilayer graphene \cite{Cao2018}. Experiments will be necessary to understand this section of the phase diagram.

\section{Acknowledgements}

The authors thank Damon Carrad, Stephen Carr, Claudio Chamon, Karsten Flensberg, Max Geier, Alex Hamilton, Thomas Jespersen, Phillip Kim, Subir Sachdev, Mathias Schuerer,  Brian Skinner, Boris Spivak, Lukas Stampfer, Oleg Sushkov, and Alexander Whiticar for helpful discussions and comments on the manuscript. TL acknowledges support from the Danish National Research Foundation and Microsoft Station Q. HS acknowledges support from the Australian--American Fulbright Commission.

\let\oldaddcontentsline\addcontentsline
\renewcommand{\addcontentsline}[3]{}

\let\addcontentsline\oldaddcontentsline

\widetext
\newpage
\begin{center}
\textbf{\large Supplemental Material}
\end{center}

\setcounter{equation}{0}
\setcounter{table}{0}
\setcounter{page}{1}
\makeatletter
\renewcommand{\theequation}{S\arabic{equation}}
\renewcommand{\bibnumfmt}[1]{[S#1]}
\renewcommand{\citenumfont}[1]{S#1}
\section*{Derivation of the Low-Energy Interacting Hamiltonian}

The generic single--particle Hamiltonian describing a superlattice with triangular symmetry is given by

\begin{gather}
H = \frac{{\bm p}^2+ p_z^2}{2m} + W_0(z) + \sum_{n\neq 0} W_n(z) \cos \left(\bm{G}_n \cdot \bm{r}\right)
\label{superlattice}
\end{gather}
where $\bm{r} = (x, y)$ are the in-plane coordinates, $z$ is the out--of--plane coordinate and $n$ indexes the reciprocal lattice vectors $\bm{G}_n$. At the $K$, $K^\prime$ points, the space of single--particle wavefunctions {$ |\Psi_{\bm k,\sigma,\tau}\rangle=e^{i\bm k\cdot \bm r} |\sigma, \tau\rangle$} is spanned by the basis of pseudospin eigenstates $|\sigma, \tau\rangle$ (with $\tau = +1, -1$ corresponding to states near the $K$ and $K^\prime$ points respectively) whose coordinate representation has the structure 
\begin{align}
\langle \bm{r},z | \sigma, \tau \rangle  &= \varphi(z) \sum_n{C_{\sigma,n} e^{i (\bm{K}_n^\tau)\cdot \bm{r}}},\\
\bm K_n^\tau &\equiv \frac{2\pi}{3L} \{ (2\tau,0), (-\tau,\sqrt{3}), (-\tau, -\sqrt{3})\}
\end{align}
The set $\{\bm K_n^{\tau}\}$ represents all Brillouin zone corners relating to a given valley $\tau$.

The many--body Hamiltonian describing  physics near the $K$, $K^\prime$ points may be expressed in terms of the fermionic creation operators $\psi^\dagger_{\bm{k},\sigma,\tau,s}$ with quasimomentum $\bm{k}$ relative to the valley momentum, pseudospin $\sigma$ and spin $s$ near the $K$ or $K^\prime$ point, in the form
\begin{gather}
H = \sum_{\bm{k}; \sigma,\sigma',\tau,s}{ \psi^\dagger_{\bm{k},\sigma,\tau,s} ( v \tau k_x \sigma^x  +v  k_y \sigma^y )_{\sigma \sigma'} \psi_{\bm{k},\sigma',\tau,s} } + \nonumber \\ \frac{1}{2} \sum_{\bm{k},\bm{p},\bm{q};\sigma_i, \tau_i, s,s'}{
U_{\sigma_1 \tau_1, \sigma_2 \tau_2, \sigma_3 \tau_3, \sigma_4 \tau_4 }(\bm{q})\psi^\dagger_{\bm{k} + \bm{q}, \sigma_3,\tau_3, s} \psi_{\bm{k}, \sigma_1, \tau_1, s} \psi^\dagger_{\bm{p}-\bm{q},\sigma_4, \tau_4, s'} \psi_{\bm{p},\sigma_2, \tau_2, s'}
} 
\end{gather}
where,
\begin{gather}
\label{bareVs}
U_{\sigma_1 \tau_1, \sigma_2 \tau_2, \sigma_3 \tau_3, \sigma_4 \tau_4 }(\bm{q}) = \sum{
	C^*_{\sigma_3 ,n'} C_{\sigma_1, n} C^*_{\sigma_4, m'} C_{\sigma_2, m} V\left(\bm{q}+ \bm{K}^{\tau_3}_{n'} - \bm{K}^{\tau_1}_n\right)
}  \nonumber \\
V(\bm{q}) = \frac{e^2}{\epsilon_r} \int{ \left[ \frac{ 1}{ \sqrt{| \bm{r}|^2 + (z- z')^2}}
	-\frac{1}{\sqrt{|\bm{r}|^2  + (z + z')^2 + 2D^2 }} \right] e^{i \bm{q}\cdot \bm{r}} \varphi(z)^2 \varphi(z')^2 dz dz' d^2\bm{r}}. \
\end{gather}
The second term in the brackets is the contribution from an image charge, resulting from a metallic gate at distance $D$ from the system. The sum in the first line is taken over values of $n, n', m, m'$ satisfying
\begin{gather}
\bm{K}^{\tau_3}_{n'} - \bm{K}^{\tau_1}_n + \bm{K}^{\tau_4}_{m'} - \bm{K}^{\tau_2}_m = 0 
\end{gather}
which can be seen to vanish unless $\tau_1 + \tau_2 = \tau_3 + \tau_4$.

The eigenfunctions of the single--particle term in the Hamiltonian are given by 
\begin{equation}
\label{DiracWf}
|\bm{k},\tau\rangle =\frac{1}{\sqrt{2}}
\left(|a,\tau\rangle+\tau e^{i\tau\theta_{ \bm{k}}}|b,\tau\rangle\right)
e^{i\bm{k}\cdot\bm{r}}
\end{equation} 
where $|a,\tau\rangle$ and $|b,\tau\rangle$ are envelope functions localized predominantly on the $A$ and $B$ sublattices, corresponding to pseudospin up and down; $\tau$ dependence is abbreviated in Eq. (10). These two components of the wavefunction are combined with a relative phase that depends on the momentum, which shifts by $\pi$ when $\bm{k}\rightarrow -\bm{k}$. As explained in the main text, this $\pi$ phase is important to the mechanism for superconductivity, as it gives rise to destructive interference in the pair wavefunction, allowing the charge densities of the two electrons to avoid being on the same sublattice.

The interaction matrix elements depend on the vertical profile of the charge density, which is sensitive to the harmonics of the electrostatic potential  $W_n(z)$. For simplicity, however, we will consider the limit of a narrow well, so the transverse wavefunctions $\varphi(z)$ are highly localized near $z = 0$ and the interaction potential $V(\bm q) $ may be replaced by 
\begin{gather}
V(\bm{q}) = \frac{e^2}{\epsilon_r} \int{ \left[\frac{1}{ |\bm{r}|} - \frac{1}{\sqrt{| \bm{r}|^2 + 4D^2}} \right] e^{i \bm{q}\cdot \bm{r}} \ d^2 \bm{r}} = \frac{2\pi e^2}{\epsilon_r |\bm{q}|} (1 - e^{-2|\bm{q}|D}) \ \ .
\end{gather}
We may ignore the term $\propto e^{-2 | \bm{q}|D}$ if the gate is far from the system, $D > 1/k_F$.

In the main text we consider a model in which the only harmonics of the superlattice potential (\ref{superlattice}) involve reciprocal lattice vectors $|\bm{G}_n| = \sqrt{3}K$ connecting points within the first Brillouin zone. We assume the potential is vertically uniform, so $W_n(z) = 2W$ is constant.

The matrix elements of the interaction vanish unless either $\tau_1 = \tau_3, \tau_2 = \tau_4$ or $\tau_1 = \tau_4 = - \tau_2 = - \tau_3$. In these cases the interactions are separable and are of the form,
\begin{align}
\notag &U=U^I+U^{II},\\
&U^I_{\sigma_1 \tau_1, \sigma_2 \tau_2, \sigma_3 \tau_3, \sigma_4 \tau_4} = \nonumber \\
&\left[V_{00;00}(\bm{q}) \sigma^0\tau^0\otimes\sigma^0\tau^0 + V_{xx;00}(\bm{q}) \sigma^x\tau^0\otimes\sigma^x\tau^0 + V_{yy;zz}(\bm{q}) \sigma^y\tau^z\otimes\sigma^y\tau^z+ V_{zz;00}(\bm{q})\sigma^z\tau^0\otimes\sigma^z\tau^0\right]_{\sigma_1 \tau_1, \sigma_3 \tau_3; \sigma_2 \tau_2, \sigma_4 \tau_4} \nonumber \\
&U^{II}_{\sigma_1 \tau_1, \sigma_2 \tau_2, \sigma_3\tau_3, \sigma_4\tau_4 } =
\left[V'_{00;\pm\mp}(\bm{q}) \sigma^0\tau^\pm\otimes\sigma^0\tau^\mp + V'_{ii;\pm\mp}(\bm{q}) \sigma^i\tau^\pm\otimes\sigma^i\tau^\mp\right]_{\sigma_1 \tau_1, \sigma_3 \tau_3; \sigma_2 \tau_2, \sigma_4 \tau_4}
\end{align}
where $i,j$ run over $x$ and $y$, 
and the indices are arranged such that, e.g. $\left[\sigma^y\tau^z\otimes\sigma^y\tau^z\right]_{\sigma_1 \tau_1, \sigma_3 \tau_3; \sigma_2 \tau_2, \sigma_4 \tau_4}\equiv (\sigma^y\tau^z)_{\sigma_1 \tau_1, \sigma_3 \tau_3}\otimes(\sigma^y\tau^z)_{\sigma_2 \tau_2, \sigma_4 \tau_4}$.

In our analysis of the superconducting state, only the matrix elements corresponding to $\tau_1 = \tau_2= \tau_3 = \tau_4$ appear, since we consider the case of Cooper pairs formed from electrons within the same valley, and hence we present only $U^I$. We take the leading order in the ratio $q\ll K$, which allows us to write
\begin{align}
V_{00;00}(\bm{q}) &= \frac{2\pi e^2}{\epsilon_r |\bm q|} +v_{00}\frac{2\pi e^2}{\epsilon_r K} \nonumber \\
V_{zz;00}(\bm{q}) &= v_{zz} \frac{2\pi e^2}{\epsilon_r K}   \nonumber \\
V_{xx;00}(\bm{q}) &= V_{yy;zz}(\bm{q}) =v_{xx} \frac{2\pi e^2}{\epsilon_r K}
\label{beta}
\end{align}
where $v_{00}, v_{zz}, v_{xx}$ are purely functions of the parameter $2m^* W/K^2$, alternatively expressed as $W/E_K$ where $E_K=K^2/(2m^*)$.

\begin{figure}[t]
	\includegraphics[width = 0.5\textwidth]{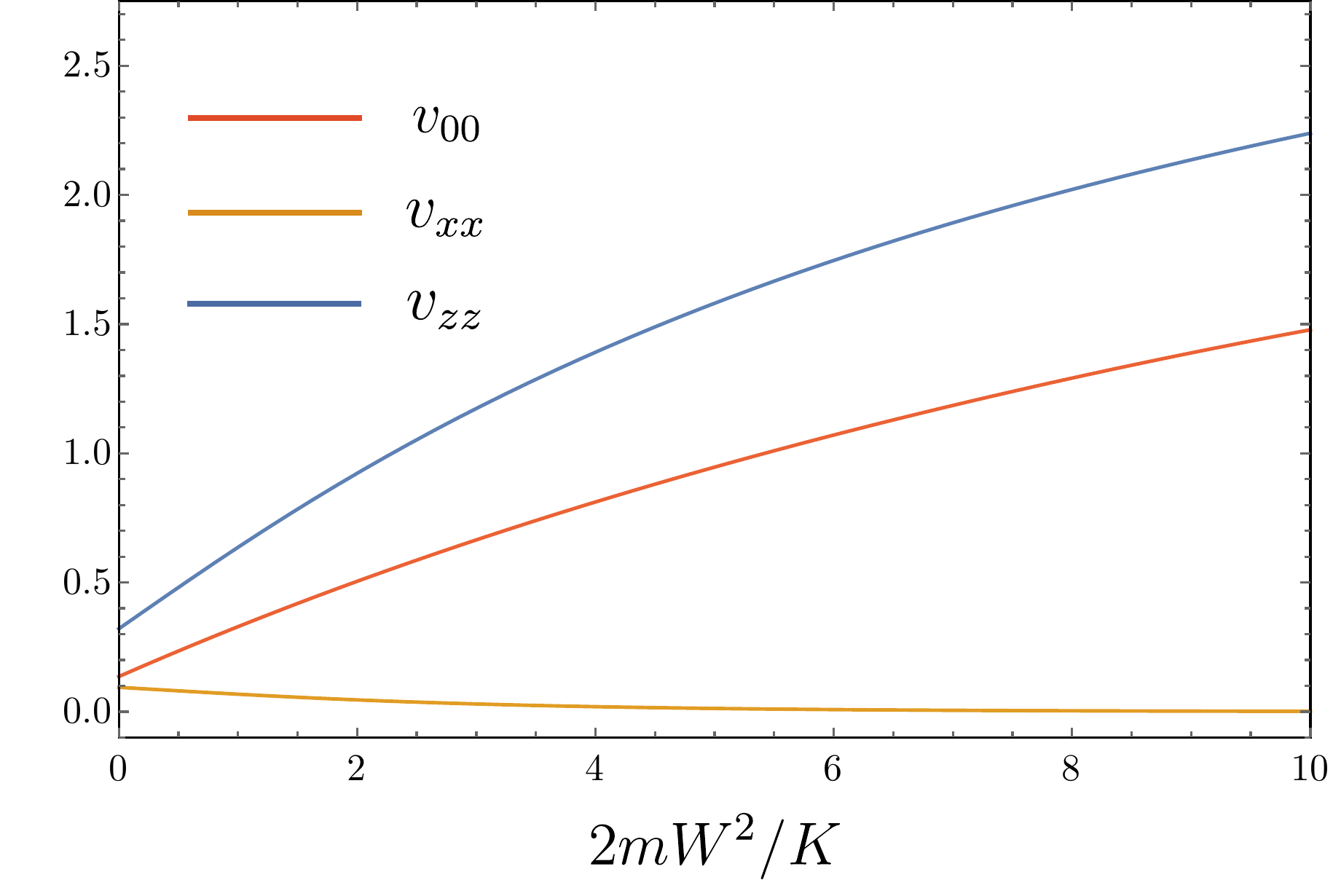} 
	\caption{The interaction constants $v_{00}, v_{xx}, v_{zz}$ defined in (\ref{beta}), as a function of the dimensionless parameter $W/E_K=2m^* W/K^2$.}
	\label{fig:couplings}
\end{figure}

These functions are plotted in Fig. \ref{fig:couplings}. In the regimes of interest, $2m^* W /K^2 >2$, we have $v_{xx} \ll v_{zz}, v_{00}$ and can be neglected. This can be understood by the fact that the off--diagonal matrix elements of $U_{\mu\nu}$ (in pseudospin indices) mix states from different sublattices. For instance, we see from Eq. \eqref{DiracWf} that $\langle \bm k',\tau' | \sigma^x|\bm k, \tau\rangle \propto \langle a,\tau'|b,\tau\rangle$.  When the atomic orbitals are strongly localized at the lattice sites, the overlap of these states is small, and hence only the diagonal terms $V_{00;00}$ and $V_{zz;00}$ remain, as reflected by the numerical results in Fig. \ref{fig:couplings}.

\newpage
\subsection{Higher Harmonics}
\begin{figure}[t]
	\center
	\hspace{-0.045\textwidth}\includegraphics[width = 0.5\textwidth]{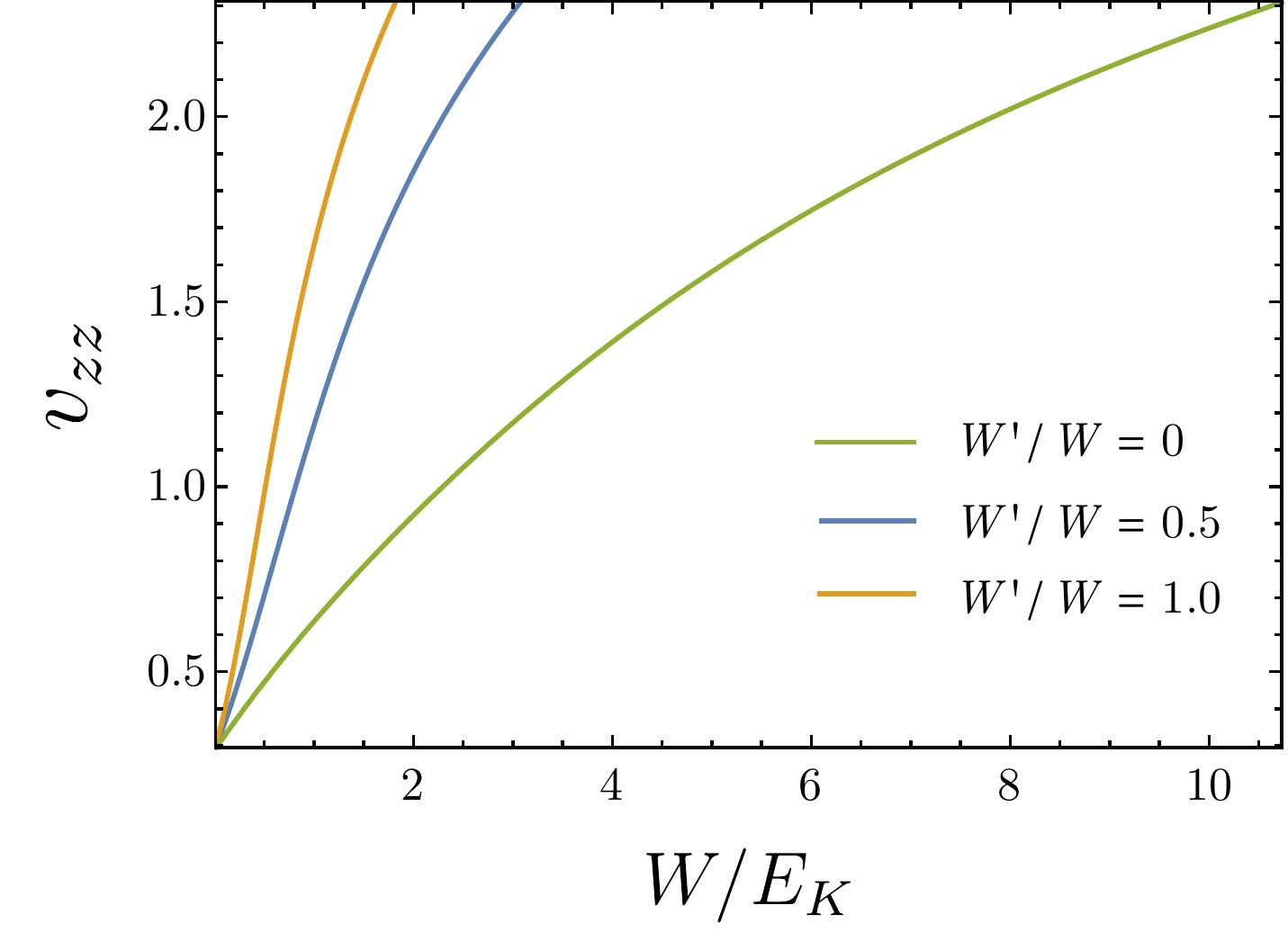} 
	\caption{The interaction constant $v_{zz}$ defined in (\ref{beta}), as a function of the dimensionless parameter $W/E_K=2m^* W/K^2$, for various values of the second harmonic potential \eqref{secondhar}, $W'/W=0,0.5,1$. }
	\label{fig:moreharmonic}
\end{figure}

While for the results in the main text we have considered the simplified potential (2), in the experimental situation terms of the form $\cos(\bm{G}\cdot \bm{r})$ will appear where $\bm{G}$ is any reciprocal lattice vector, and terms in which $|\bm{G}| > \sqrt{3}K$ may be comparable to the terms previously considered. It is therefore useful to consider the effects of possible higher harmonics on the strengths of the interactions. Experimentally, these additional harmonics become more important when the ratio of the antidot size to the unit cell is increased, and therefore the electron wavefunctions are localized more effectively. In Figure \ref{fig:moreharmonic}, results are presented for $v_{zz}$ for a model superlattice potential with an additional set of harmonics

\begin{align}
U'(\bm{r})  \ = \  W \sum_{|\bm{G}| = \sqrt{3}K}{ \cos \left(\bm{G}\cdot \bm{r}\right)} - W' \sum_{|\bm{G}'| = 3K} { \cos\left( \bm{G}'\cdot \bm{r}\right)} 
\label{secondhar}
\end{align}

The figures show the dimensionless interaction $v_{zz}$ as a function of $W/E_K$ for fixed ratios $W'/W = 0, 0.5, 1$. We see that the additional harmonics greatly reduce the value of $W$ required to achieve the superconducting state. Note that the additional harmonics $\propto W'$ do not affect the total variation of the potential.

\newpage
\section*{Screened Interactions}

The RPA equations for the screened interactions are
\begin{align}
\widetilde{V}_{\mu \nu; \rho \lambda}(\omega, \bm{q}) = 
 V_{\mu \nu; \rho \lambda}(\bm{q}) + V_{\mu \alpha; \rho \beta}(\bm{q}) \Pi^{\alpha \gamma; \beta \delta} (\omega,\bm{q}) \widetilde{V}_{\gamma \nu; \delta \lambda} (\omega, \bm{q})
\end{align} where $\Pi^{\alpha \gamma; \beta \delta}$ is the polarization operator, 

\begin{gather}
\Pi^{\alpha \gamma; \beta \delta}(\omega, \bm{q}) = 
-i \text{Tr} \int{
\sigma^\alpha \tau^\beta G(E+\omega, \bm{k}+\bm{q}) \sigma^\gamma \tau^\delta G(E, \bm{k}) \frac{ dE d^2\bm{k}}{(2\pi)^3}
} \ \ , \nonumber \\
G(E, \bm{k}) = \frac{1}{ E +\mu - v ( \tau^z k_x \sigma^x + k_y \sigma^y) +i 0 \text{sgn}(E)}
\label{Pi}
\end{gather} 
The RPA equations for intravalley and intervalley scattering decouple, and in this paper we restricted our discussion to the intravalley interactions, ie $\rho=\lambda = 0,z$. From the results in the previous section, we may set $V_{\mu \nu;\rho\lambda}(\bm{q}) \rightarrow 0$ unless $\mu=\nu=0,z$ and $\rho=\lambda=0$. In this case, the solution to the RPA equations has the simple form, where
\begin{align}
\label{RpaSols}
\widetilde{V}_{00;00}(\omega,\bm q)  = \frac{ V_{00;00}(\bm{q})}{1 - \Pi^{00;00}(\omega,\bm q) V_{00;00}(\bm{q})},  \ \ \widetilde{V}_{zz;00}(\omega,\bm q) = \frac{ V_{zz;00}(\bm{q})}{1 - \Pi^{zz;00}(\omega,\bm q) V_{zz;00}(\bm{q})} 
\end{align}   
and all other interactions are zero.

The polarization operators $\Pi^{\alpha\gamma;\beta\delta}$ can be calculated using dimensional regularization, using similar manipulations to those used to calculate $\Pi^{00;00}$ for graphene, c.f. \cite{Son2007, Wunsch2006}. Since the calculation of $\Pi^{00;00}$ is presented many papers on graphene, here we will give the derivation of polarization operator $\Pi^{zz;00}$, which as far as we are aware has not been previously discussed.

Inserting the expression for the electron Green's function, the formula for the polarization operator gives
\begin{align}
\label{PolOp1}
\Pi^{\alpha\gamma;\beta\delta}(\omega,\bm q)= -i\text{Tr} \int \frac{d^2k}{(2\pi)^2}\frac{dE}{2\pi}\frac{J^{\alpha\beta}((\omega+E)e^{i0}+\mu+(\bm k+\bm q)\cdot \widetilde{\bm \sigma})J^{\gamma\delta}(E e^{i0}+\mu+\bm k\cdot \widetilde{\bm \sigma})}{(((\omega+E)e^{i0}+\mu)^2-(\bm k+\bm q)^2)((E e^{i0}+\mu)^2-k^2)}
\end{align}
where $\widetilde{\bm \sigma}=(\tau^z\sigma^x,\sigma^y)$, $J^{\alpha\beta}=\sigma^\alpha \tau^\beta$ and $J^{\gamma\delta}=\sigma^\gamma \tau^\delta$, and we will focus on $J^{\alpha\beta}=\sigma^z\tau^0$, $J^{\gamma\delta}=\sigma^z\tau^0$. For brevity, we will also abbreviate $\mu=(\alpha,\beta)$, $\nu=(\gamma,\delta)$. We will work in units where the Dirac velocity is unity. The replacement $e^{i0} = 1+ i0$ makes the notation more compact, and since $E e^{i0} = E(1+i0) = E + i0 \text{sgn} E$ we see that it is equivalent to the more common notation for the $i0$ prescription and the one used in the main text Eq. (7).

The poles of the integrand in the frequency plane are at $E = -\omega +(-\mu\pm |\bm k+\bm q|)e^{-i0}, (-\mu\pm k)e^{-i0}$, two of which are always located in the upper half plane, the other two are located in the upper half plane if $\mu - k >0$, or $\mu - |\bm k-\bm q| >0$, respectively. Closing the integration contour in the upper half plane and performing the frequency integral by residues, we obtain four contributions, one from each pole, 
\begin{align}
\Pi^{\mu\nu}(\omega,q)= \int \frac{d^2k}{(2\pi)^2}  \frac{J^\mu (\tilde{\omega}-k + (\bm k + \bm q )\cdot \widetilde{\bm \sigma})J^\nu(-k+ \bm k \cdot \widetilde{\bm \sigma})}{-2k(\tilde{\omega}-k+|\bm k + \bm q|)(\tilde \omega-k-|\bm k + \bm q| )} + 
\frac{J^\mu (-|\bm k + \bm q| + (\bm k + \bm q )\cdot \widetilde{\bm \sigma})J^\nu(-\tilde \omega-|\bm k + \bm q|+ \bm k \cdot \widetilde{\bm \sigma})}{-2|\bm k + \bm q |(-\tilde{\omega}-|\bm k + \bm q|+k)(-\tilde \omega-|\bm k + \bm q|-k )} \nonumber\\
\frac{J^\mu (\tilde{\omega}+k +(\bm k + \bm q )\cdot \widetilde{\bm \sigma})J^\nu(k+ \bm k \cdot \widetilde{\bm \sigma})}{-2k(\tilde{\omega}-k+|\bm k + \bm q|)(\tilde \omega-k-|\bm k + \bm q| )} \Theta(\mu-k) + 
\frac{J^\mu (|\bm k + \bm q| + (\bm k + \bm q )\cdot \widetilde{\bm \sigma})J^\nu(-\tilde \omega+|\bm k + \bm q|+\bm k \cdot \widetilde{\bm \sigma})}{2|\bm k + \bm q |(-\tilde{\omega}+|\bm k + \bm q|-k)(-\tilde \omega+|\bm k + \bm q|+k )}  \Theta(\mu-|\bm k + \bm q|)
\end{align}  
For compactness, we have used the notation $\tilde{\omega} = \omega(1+i0)$. Substituting $\bm k \rightarrow \bm k -\bm q$ in the second and fourth contributions, and combining terms, we get
\begin{align}
\label{freqint}
\Pi^{\mu\nu}(\omega,q) =\sum_{s=\pm 1} \int \frac{d^2k}{(2\pi)^2} \frac{ A_s^{\mu\nu} - B_s^{\mu\nu} \Theta(\mu-k)}{2k( (s\tilde \omega - k)^2 -(sq+k)^2 )}
\end{align}
 where 
 \begin{align}
A_s^{\mu\nu} &= \text{Tr} \ J^{s_1}(s\tilde{\omega}+k-(\bm k - s\bm q)\cdot \widetilde{\bm{\sigma}})J^{s_2}(-k+\bm{k} \cdot \widetilde{\bm \sigma})\\
\label{intranum}
B_s^{\mu\nu} &= \text{Tr} \ J^{s_1}(s\tilde{\omega}+k+(\bm k + s\bm q)\cdot \widetilde{\bm \sigma})J^{s_2}(k+\bm k \cdot \widetilde{\bm \sigma})
 \end{align}
with $(s_1,s_2)=(\mu,\nu)$ when $s=1$ and $s_1\leftrightarrow s_2$ when $s=-1$. 
 
The first term in \eqref{freqint}, which contributes when $\mu=0$, is the so--called `interband' polarization operator, denoted $\Pi_-^{\mu\nu}$. The second, which vanishes unless $\mu \neq 0$, is the `intraband' polarization operator $\Pi_+^{\mu\nu}$. The former is linearly divergent and requires regularization, while the second is manifestly finite due to the theta function. We compute them separately. We will take the external frequency to be positive $\omega>0$.
 
The interband part is most straightforwardly calculated by returning to the original expression Eq. \eqref{PolOp1} and setting $\mu=0$. We remove divergences using dimensional regularization, using the formulae
\begin{gather}
\label{dimreg1}
\int \frac{d^dk}{(2\pi)^d} \frac{(k^2)^a}{(k^2+\Delta)^b} = \frac{1}{(4\pi)^{d/2}}\ \frac{\Gamma(\frac{d}{2}+a)\Gamma(b-a-\frac{d}{2})}{\Gamma(\frac{d}{2})\Gamma(b)}\ \Delta^{d/2+a-b} \\
\label{dimreg2}
\int \frac{d^dk}{(2\pi)^d} \frac{(k^2)^a}{(k^2-\Delta)^b} = i\ \frac{(-1)^{a-b}}{(4\pi)^{d/2}}\ \frac{\Gamma(\frac{d}{2}+a)\Gamma(b-a-\frac{d}{2})}{\Gamma(\frac{d}{2})\Gamma(b)}\ \Delta^{d/2+a-b}
\end{gather}
Focus first on the denominator in Eq. \eqref{PolOp1}. Using the Schwinger--Feynman parametrization,
\begin{align}
\frac{1}{AB} = \int_0^1\frac{dx}{(xA+(1-x)B)^2}
\end{align}
we write
\begin{align}
\frac{1}{((\omega+E)^2-(\bm{k}+\bm{q})^2)(E^2-k^2)}=\int_0^1 dx \ \frac{1}{((k+xq)^2-x(x-1)q^2)^2}
\end{align}
where we now use relativistic notation $l^\mu=(E,\bm{k})$, $p^\mu=(\omega,\bm{q})$, and $l^2=E^2-\bm{k}^2$, $p^2=\omega^2-\bm{q}^2$. Shifting $l\rightarrow l-xp$ (bear in mind that this will affect the numerator as well), and Wick rotating $E \rightarrow iE$, the expression becomes
\begin{align}
\int_0^1 dx \ \frac{1}{(l^2+x(x-1)p^2)^2}
\end{align}
The corresponding numerator (before Wick rotation) is 
\begin{align}
\text{Tr} \  J^\mu \left((1-x)\omega+E + [(1-x)\bm q+ \bm k]\cdot \widetilde{\bm \sigma} \right) J^\nu \left( -x\omega + E +[-x\bm q+\bm k]\cdot\widetilde{\bm \sigma}\right)
\end{align}
Now we substitute $J^\mu=J^\nu=\sigma^z \otimes \tau^0$ and perform the trace over pseudospin, spin and valley indices (note that the trace of the identity is therefore 8). Throwing away all terms in the numerator which are linear in $l^\mu$, since these integrate to zero, we obtain the expression 
 \begin{align}
 \Pi^{zz;00}_- = \int_0^1 dx  \int \frac{d^dl}{(2\pi)^d} \frac{8x(x-1)p^2+ 8l^2 }{(k^2+\Delta)^2}
 \end{align}
with $\Delta=x(x-1)p^2$. Using \eqref{dimreg1} we find the real part,
 \begin{align}
 \text{Re} \ \Pi^{zz;00}_- = \int_0^1 dx \  \left\{ \frac{1}{8\pi}\frac{8x(1-x)p^2}{\sqrt{x(x-1)p^2}}+\frac{3}{8\pi}\cdot 8\sqrt{x(x-1)p^2} \right\}\Theta(-p^2)
 \end{align}
Using $\int_0^1 dx \sqrt{x(1-x)} = \pi/8$, we arrive at 
 \begin{align}
 \text{Re} \ \Pi^{zz;00}_- = \frac{1}{2}\sqrt{q^2-\omega^2}\ \Theta(q-\omega)
 \end{align}

We note that there is a subtlety if one tries to calculate the imaginary part from the real part using the Kramers--Kronig relations, which arises due to the fact that $\Pi^{zz;00}$ does not drop off $\lesssim 1/\omega$ at high frequencies, resulting in a contribution from the contour at infinity. In such situations, the correct dispersion relations require subtractions to account for this additional contribution. Through this approach or through using \eqref{dimreg2} to calculate the imaginary part directly, one finds 
 \begin{align}
 \text{Im} \ \Pi^{zz;00}_- = -\frac{1}{2}\sqrt{\omega^2- q^2}\ \Theta(\omega-q)
 \end{align}

We now turn to the intraband contributions. We will work with the form Eq. \eqref{freqint}. Firstly consider the integral,
\begin{align}
\label{cos1}
I(a)=\int_0^{2\pi} \frac{d\theta}{a+\cos\theta} = \frac{\text{sgn}(a)}{\sqrt{a^2-1}} \Theta(|a|-1)
\end{align}
This result may be derived by the substitution $z=e^{i\theta}$, whereupon the integral over $\theta$ becomes an integral around the unit circle,
\begin{align}
\label{contour}
\int_0^{2\pi} \frac{d\theta}{a+\cos\theta} =\oint \frac{dz}{2\pi i}\frac{1}{z^2+2az + 1} = \oint \frac{dz}{2\pi i}\frac{1}{(z-z_1)(z-z_2)}
\end{align}
where $z_{1,2}=-a\pm \sqrt{a^2-1}$. Note that since $z_1z_2=1$, if one root is outside the unit circle the other is inside. In the case of real $a$, when $a<1$ then both roots sit exactly on the unit circle. If both poles are included in the integration, their contributions cancel and the integral vanishes. When $z_i$ is inside the circle and $z_j$ is outside, the integral equals
\begin{align}
\oint \frac{dz}{2\pi i}\frac{1}{(z-z_1)(z-z_2)} = \frac{1}{z_i-z_j}
\end{align}
which can straightforwardly be shown to give Eq. \eqref{cos1}.

Now consider the case where $a$ is given an infinitesimal imaginary part $a\rightarrow a+ ib 0  $.  Returning to \eqref{contour}, the poles are now located at $z_{1,2}=-a-ib0  \pm \sqrt{(a+ib0)^2-1}$. The integral is unaffected by the shift in the poles in the case $a>1$, however when $a<1$, one of the poles is shifted just outside the unit circle, and there is no cancellation between the two poles. Hence the integral obtains a nonzero contribution for $a>1$, which results in
\begin{align}
\label{cos2}
 I(a + ib0)=\int_0^{2\pi} \frac{d\theta}{a+ib0+\cos\theta} = \frac{\text{sgn}(a)}{\sqrt{a^2-1}} \Theta(|a|-1)-i \frac{\text{sgn}(b)}{\sqrt{1-a^2}}\ \Theta(1-|a|)
\end{align}
These integrals will be used to calculate the real and imaginary parts of the intraband polarization operator.

 Performing the trace in \eqref{intranum} with $J^\mu=J^\nu=\sigma^z \otimes \tau^0$, we obtain 
 \begin{align}
 \label{interbandint}
\int &\frac{d^2k}{k(2\pi)^2}\sum_{s=\pm} \  \frac{4s\tilde{\omega}k-4s\bm k \cdot \bm q}{\tilde{\omega}^2+2s\tilde{\omega}k-q^2-2s\bm k \cdot \bm q}  \Theta\left(\mu-k \right) = \frac{2\mu}{\pi}+\int \frac{d^2k}{k(2\pi)^2}\sum_{s=\pm} \ \frac{1}{2skq}\frac{2(\omega^2-q^2)}{-\frac{\tilde{\omega}^2+2s\tilde{\omega}k-q^2}{2skq}+\cos\theta}\Theta\left(\mu-k \right) 
 \end{align}

Note the constant term, given by $N \times \mu/(2\pi)$ where $N=4$, is positive. By contrast, calculation of the density--density response $\Pi^{00;00}$ finds a similar expression, but with a constant term $N \times -\mu/(2\pi)$. The relative minus sign arose in the calculation of $\Pi^{zz;00}$ from anticommuting $\sigma^z$ past the single particle Hamiltonian $\propto \bm k \cdot \widetilde{\bm \sigma}$. Since the second term in this expression vanishes at small $q$ and $\omega$, in the long--wavelength and static limit the functions $\Pi^{zz;00}$ and $\Pi^{00;00}$ equal a positive and negative constant respectively. The negative value of $\Pi^{00;00}$ is responsible for screening ie weakening of charge density fluctuations $\widetilde{V}_{00;00}<{V}_{00;00}$, while the positive value of $\Pi^{zz;00}$ leads to antiscreening ie strengthening of pseudospin fluctuations $\widetilde{V}_{zz;00}>{V}_{zz;00}$, as can be seen from \eqref{RpaSols}. It is precisely these antiscreened interactions which give rise to intravalley pairing, and so as the efficiency of screening is increased -- for instance through doping -- so does the tendency toward superconductivity.

We now take the real part of \eqref{interbandint} using \eqref{cos2}. Performing the angular integral, 
 \begin{align}
\text{Re} \ \Pi^{zz;00}_+ = \frac{2\mu}{\pi}+\int \frac{dk}{2\pi} \frac{2(q^2-\omega^2) }{ \sqrt{ (\omega^2+2s\tilde{\omega}k-q^2)^2-4k^2q^2} } \ \text{sgn}(\omega^2+2s{\omega}k-q^2 ) \  \Theta( (\omega^2+2s{\omega}k-q^2)^2-4k^2q^2 ) \ \Theta\left(\mu-k \right) 
\end{align}
Changing variables $k\rightarrow k-s\omega/2$ simplifies this expression to 
\begin{align}
\text{Re} \ \Pi^{zz;00}_+=\frac{2\mu}{\pi}+\int \frac{dk}{2\pi} \frac{2(q^2-\omega^2)}{\sqrt{(\omega^2-q^2)(4k^2-q^2)}}\ \text{sgn}( 2s\omega k - q^2 ) \ \Theta((\omega^2-q^2)(4k^2-q^2) )\ \Theta\left(\mu+s\tfrac{\omega}{2}-k \right)\ \Theta\left(k-s\tfrac{\omega}{2} \right)
\end{align}
Since $k$ originally ranged from 0 to $\infty$, after the change of variables $k$ ranges from $-s\omega/2$ to $\infty$. To make this clear, we added an additional theta function in the previous expression emphasizing the correct limits of integration. To proceed further, we will require the integrals,
\begin{align}
\label{Omegazz1}
 \int \frac{dk}{\sqrt{q^2-4k^2}} &= \frac{1}{2}\sin^{-1}\left( \frac{2k}{q} \right) \\
\label{Omegazz2}
\int \frac{dk}{\sqrt{4k^2-q^2}} &= \frac{1}{2} \log \left( 2k + \sqrt{4k^2-q^2} \right)
\end{align}

We have to consider the cases (a) $\omega > q$ and (b) $\omega < q$ separately. Firstly for case (a), we have 
 \begin{align}
\text{Re} \ \Pi^{zz;00}_+= \frac{2\mu}{\pi} -2\sqrt{\omega^2-q^2}\int \frac{dk}{2\pi}  \frac{1}{\sqrt{4k^2-q^2}}\ \text{sgn}\left( 2s\omega k - q^2 \right) \Theta\left(4k^2-q^2 \right)\Theta\left(\mu+s\tfrac{\omega}{2}-k \right)\Theta\left(k-s\tfrac{\omega}{2} \right)
\end{align}
Paying careful attention to the limits imposed by the theta functions, and making use of \eqref{Omegazz2}, the integral results in 
\begin{align}
\text{Re} \ \Pi^{zz;00}_+= \frac{2\mu}{\pi} -\frac{1}{2\pi}\sqrt{\omega^2-q^2} \left( \log\left( \tfrac{2\mu + \omega + \sqrt{(2\mu+\omega)^2-q^2}}{\omega+ \sqrt{\omega^2-q^2}} \right) \right.&+ \text{sgn} \left(\omega-2\mu+q \right)\log\left( \tfrac{|2\mu - \omega + \sqrt{(2\mu-\omega)^2-q^2}|}{\omega+ \sqrt{\omega^2-q^2}} \right)\Theta\left(|2\mu - \omega|-q \right) ) \nonumber \\
- \log\left(\tfrac{|\omega-\sqrt{\omega^2-q^2}|}{q} \right) &\left.\left\{\Theta\left(2\mu - \omega+ q \right)+\Theta\left(2\mu - \omega- q \right) \right\} \right)
\end{align}

Turning to case (b), we now have  
 \begin{align}
\text{Re} \ \Pi^{zz;00}_+= \frac{2\mu}{\pi}+2\sqrt{q^2-\omega^2}\int \frac{dk}{2\pi}  \frac{1}{\sqrt{q^2-4k^2}}\ \text{sgn}\left( 2s\omega k - q^2 \right) \Theta\left(q^2-4k^2 \right)\Theta\left(\mu+s\tfrac{\omega}{2}-k \right)\Theta\left(k-s\tfrac{\omega}{2} \right)
\end{align}
Making use of \eqref{Omegazz1}, this integral equals 
 \begin{align}
\text{Re} \ \Pi^{zz;00}_+=\frac{2\mu}{\pi}-\frac{1}{2\pi}\sqrt{q^2-\omega^2}\left(\sin^{-1}\left(\tfrac{2\mu+ \omega}{q} \right) \Theta\left(q-|2\mu+\omega| \right) \right. & \left.+\sin^{-1}\left(\tfrac{2\mu- \omega}{q} \right) \Theta\left(q-|2\mu-\omega| \right) \right. \nonumber \\
&+ \left. \frac{\pi}{2} \left\{\Theta \left( |2\mu + \omega| - q  \right)+\Theta \left( |2\mu - \omega| - q  \right)\right\} \right)
\end{align}  

To calculate the imaginary part, we use \eqref{cos2}, observing that the imaginary part of the denominator in \eqref{interbandint} is $\propto -(\omega + sk)\times 0$. Hence, following the same simplifications we applied to the real part,
\begin{align}
\text{Im} \ \Pi^{zz;00}_+ = -\int \frac{dk}{2\pi} \frac{2(q^2- \omega^2)}{\sqrt{(\omega^2-q^2)(4k^2-q^2)}}\text{sgn}\left( \omega + sk \right) \Theta\left( (\omega^2-q^2)(q^2-4k^2) \right)\Theta\left(\mu+s\tfrac{\omega}{2}-k \right)\Theta\left(k-s\tfrac{\omega}{2} \right)
\end{align}

Once again we must separately consider cases (a) $\omega > q$ and (b) $\omega < q$. Through the same manipulations which result in the real part, case (a) results in 
\begin{align}
\text{Im} \ \Pi^{zz;00}_+ =-\frac{1}{2\pi}\sqrt{\omega^2-q^2} \left( \sin^{-1}\left(\tfrac{|2\mu-\omega|}{q} \right) \Theta \left(q-|2\mu-\omega| \right) +\frac{\pi}{2}\left\{ \Theta\left( 2\mu- \omega + q \right)+\Theta\left( 2\mu- \omega - q \right) \right\} \right)
\end{align}
while case (b) results in 
\begin{align}
\text{Im} \ \Pi^{zz;00}_+ =\frac{1}{2\pi}\sqrt{q^2-\omega^2}\left( \log\left(\tfrac{2\mu + \omega + \sqrt{(2\mu+\omega)^2-q^2}}{q} \right)\Theta\left(2\mu + \omega - q \right)- \log\left(\tfrac{2\mu - \omega + \sqrt{(2\mu-\omega)^2-q^2}}{q} \right)\Theta\left(2\mu - \omega - q \right)  \right)
\end{align}

The addition of the inter and intraband polarization operators give the total expression for $\Pi^{zz;00}$. The above results can be summarized as
\begin{align}
\label{PolOpZResult}
\Pi^{zz;00}(\omega,\bm q)&= \Pi^{zz;00}_+(\omega,\bm q) + \Pi^{zz;00}_-(\omega,\bm q)
\end{align}
where 
\begin{align}
\Pi^{zz;00}_-(\omega,\bm q)= \frac{1}{2}\sqrt{\omega^2-q^2} \ \Theta\left(\omega - q \right)-i\frac{1}{2}\sqrt{q^2-\omega^2} \ \Theta\left(q -\omega\right)
\end{align}
and
\begin{align}
\Pi^{zz;00}_+(\omega,\bm q) = \Pi^{zz;00}_{+,1}(\bm{q},\omega) \ \Theta\left(\omega-q\right)+\Pi^{zz;00}_{+,2}(\bm{q},\omega) \ \Theta\left(q-\omega\right)
\end{align}
with real and imaginary parts 
\begin{align} 
\text{Re} \ \Pi^{zz;00}_{+,1}(\omega,\bm q)  &= \frac{2\mu}{\pi}- \frac{1}{2\pi}\sqrt{\omega^2-q^2} \left(f_1^{(z)}(\omega,q)\Theta(|2\mu+\omega|-q) + \text{sgn}\left(\omega-2\mu+q \right)f_1^{(z)}(-\omega,q)\Theta\left(|2\mu-\omega|-q \right) \right. \nonumber \\
& \ \ \ \ \ \ \ \ \ \ \ \ \ \ \ \ \ \ \ \ \ \ \ \ \ \ \ \ \ \ \ \ \ \ \ \ \ \ \ \left.-f_2^{(z)}(\omega,\bm q)\left\{\Theta\left(2\mu - \omega+ q \right)+\Theta\left(2\mu - \omega- q \right) \right\}\right)\\
\text{Im} \ \Pi^{zz;00}_{+,1}(\omega,\bm q)  &= -\frac{1}{2\pi}\sqrt{q^2-\omega^2} \  \left( f_3^{(z)}(-\omega,q) \Theta \left(q-|2\mu-\omega| \right)+\frac{\pi}{2}\left\{\Theta\left( 2\mu+ \omega - q \right)+\Theta\left( 2\mu- \omega - q \right)\right\} \right)\\
\text{Re} \ \Pi^{zz;00}_{+,2}(\omega,\bm q) &= \frac{2\mu}{\pi}- \frac{1}{2\pi}\sqrt{q^2-\omega^2}\left(f_3^{(z)}(\omega,q)\Theta\left(q-|2\mu+\omega| \right)+f_3^{(z)}(-\omega,q)\Theta\left(q-|2\mu-\omega| \right) \right. \nonumber \\
& \ \ \ \ \ \ \ \ \ \ \ \ \ \ \ \ \ \ \ \ \ \ \ \ \ \ \ \ \ \ \ \ \ \ \ \ \ \ \ \ \ \ \ \ \ \ \ \ \ \ \ \ \ \ \ \ \ \ \ \  + \left.\frac{\pi}{2}\left\{\Theta\left(|2\mu-\omega| -q\right)+ \Theta\left(|2\mu+\omega| -q\right)\right\} \right)\\
\text{Im} \ \Pi^{zz;00}_{+,2}(\omega,\bm q)&= \frac{1}{2\pi}\sqrt{q^2-\omega^2}\left(f_4^{(z)}(\omega,q)\Theta\left(2\mu+\omega-q \right)-f_4^{(z)}(-\omega,q)\Theta\left(2\mu-\omega-q \right)  \right)
\end{align}
where
\begin{align}
\notag f^{(z)}_1(\nu,\bm q) &= \log\left(\tfrac{|2\mu+\nu+ \sqrt{(2\mu+\nu)^2-q^2}|}{\nu + \sqrt{\nu^2-q^2}} \right)\\
\notag f^{(z)}_2(\nu,\bm q)&= \log \left( \tfrac{ |\nu -\sqrt{\omega^2-q^2} |}{q}\right)\\
\notag f^{(z)}_3(\nu,\bm q) &= \sin^{-1}\left(\tfrac{2\mu+\nu}{q} \right)\\
f^{(z)}_4(\nu,\bm q) &= \log\left(\tfrac{|2\mu+\nu+ \sqrt{(2\mu+\nu)^2-q^2}|}{q} \right)
\end{align}

To make the physics of antiscreening more transparent, we observe that the real part for $\omega<q$ is 
\begin{align}
\label{PolOpZReal}
\text{Re} \ \Pi^{zz;00}(\omega,\bm q)&= \frac{1}{2}\sqrt{q^2 - \omega^2}+ \frac{2\mu}{\pi } - \frac{1}{2\pi} \sqrt{q^2 - \omega^2}\sum_i{ \sin^{-1} \tfrac{ p_i}{q}}
\end{align}
where $p_i = \{\text{min}( 2\mu+ \omega, q),\text{min}(2\mu - \omega, q)\}$, and that this expression for $\omega=0$ and $q<2\mu$ is simply
\begin{align}
\text{Re} \ \Pi^{zz;00}(\omega=0,q<2\mu)  = \frac{2\mu}{\pi }
\end{align} 
as stated in the main text Eq. (9). Since $\Pi^{zz;00}>0$, the denominator in \eqref{RpaSols}, given by $1-\Pi^{zz;00}V_{zz;00}<1$, and hence $V_{zz;00}$ is enhanced.

For completeness, we also state without proof the known expression for $\Pi^{00;00}$, which has a similar form. The interband polarization operator is
\begin{align}
\Pi^{00;00}_-(\omega,\bm q) = -\frac{ q^2 }{4\sqrt{\omega^2-q^2}} \ \Theta\left(\omega - q \right)-i\frac{  q^2 }{4\sqrt{q^2-\omega^2}}\ \Theta\left(q -\omega\right)
\end{align}
while the intraband polarization operator is given by 
\begin{align}
\label{PolOp0Result}
\Pi^{00;00}_+(\omega,\bm q) = \Pi^{00;00}_{+,1}(\bm{q},\omega) \ \Theta\left(\omega-q\right)+\Pi^{00;00}_{+,2}(\bm{q},\omega) \ \Theta\left(q-\omega\right)
\end{align}
where 
\begin{align}
\text{Re} \ \Pi^{00;00}_{+,1}(\omega,\bm q)&= -\frac{2\mu}{\pi}+ \frac{1}{4\pi\sqrt{\omega^2-q^2} }\left( f^{(0)}_1(\omega,q)\ \Theta(|2\mu +\omega|-q) + \text{sgn}(\omega-2\mu+q)f_1^{(0)}(-\omega,q)\ \Theta(|2\mu -\omega|-q) \right. \nonumber \\
& \ \ \ \ \ \ \ \ \ \ \ \  \ \ \ \ \ \ \ \ \ \ +  \left. f_2^{(0)}(\omega,q)\left\{\Theta(2\mu-\omega+q)+\Theta(2\mu-\omega-q) \right\} \right)\\
\text{Im} \ \Pi^{00;00}_{+,1}(\omega,\bm q) &= \frac{1}{4\pi\sqrt{\omega^2-q^2} }\ \left( f_3^{(0)}(-\omega,q)\ \Theta(q-|2\mu-\omega|) + \frac{\pi q^2}{2}\left\{ \Theta(2\mu - \omega+q)+\Theta(2\mu - \omega-q) \right\}\right) \\
\text{Re} \ \Pi^{00;00}_{+,2}(\omega,\bm q) &= -\frac{2\mu}{\pi}+\frac{1}{4\pi\sqrt{q^2-\omega^2} }\left( f^{(0)}_3(\omega,q)\ \Theta\left(q-|2\mu + \omega| \right) + f^{(0)}_3(-\omega,q)\ \Theta\left(q-|2\mu - \omega| \right) \right. \nonumber \\
& \ \ \ \ \ \ \ \ \ \ \ \  \ \ \ \ \ \ \ \ \ \ + \left. \frac{\pi q^2}{2}\left\{ \Theta(|2\mu + \omega|+q)+\Theta(|2\mu - \omega|-q) \right\} \right)\\
\text{Im} \ \Pi^{00;00}_{+,2}(\omega,\bm q) &= -\frac{1}{4\pi\sqrt{q^2-\omega^2} }\Theta\left(2\mu+\omega-q \right) \ \left(f_4^{(0)}(\omega,q) - f_4^{(0)}(-\omega,q)\ \Theta(2\mu-\omega-q) \right) 
\end{align}
where 
\begin{align}
\label{PolOp0Result}
\notag f^{(0)}_1(\nu,\bm q) &= (2\mu+\nu)\sqrt{(2\mu+\nu)^2-q^2}-q^2 \log\left(\tfrac{|2\mu+\nu +\sqrt{(2\mu+\nu)^2-q^2}|}{|\nu + \sqrt{\nu^2-q^2}|} \right)\\
\notag f^{(0)}_2(\nu,\bm q)&= q^2 \log \left( \tfrac{ |\nu -\sqrt{\nu^2-q^2}| }{q}\right)\\
\notag f^{(0)}_3(\nu,\bm q) &= (2\mu+\nu)\sqrt{q^2-(2\mu+\nu)^2}+q^2 \sin^{-1}\left(\tfrac{2\mu+\nu}{q} \right)\\
f^{(0)}_4(\nu,\bm q) &= (2\mu+\nu)\sqrt{(2\mu+\nu)^2-q^2}-q^2 \log\left(\tfrac{|2\mu+\nu +\sqrt{(2\mu+\nu)^2-q^2}|}{q} \right)
\end{align}
in accordance with Ref. \cite{Hwang2007b}, noting that we have assumed $\omega>0$, and that the definition of $\Pi$ in Ref. \cite{Hwang2007b} differs from the conventional one by a relative minus sign, as is clear from their Eq. (2). The derivation of this expression is similar to the one we presented for $\Pi^{zz;00}$, but with $J^\mu=J^\nu =\sigma^0\tau^0$ and therefore a different trace structure in the numerator of \eqref{freqint}.

Again, the screening effects of this function can be made manifest by observing that the real part of this expression for $\omega<q$ is simply 
\begin{align}
\label{PolOp0Result2}
\Pi^{00;00}(\omega,\bm q) &= - \frac{q^2}{4 \sqrt{q^2 - \omega^2}} - \frac{ 2 \mu}{\pi } + \frac{1}{4\pi \sqrt{q^2 - \omega^2}} \sum_i{
p_i{ \sqrt{q^2 - p_i^2}} + q^2 \sin^{-1} \tfrac{p_i}{q}} 
\end{align}
which in the limit of $\omega=0$ and $q<2\mu$ is 
\begin{align}
\text{Re} \ \Pi^{00;00}(\omega=0,q<2\mu) = -\frac{2\mu}{\pi }
\end{align} 
corresponding to conventional screening ie weakening of the effective value of $V_{00;00}$.

The derivation of the polarization operator for intervalley scattering follows the same kind of manipulations as those above, but our calculations found the intravalley  interactions to be the dominant effect, so we omit these results.

\section*{Zero Temperature Gap}
The gap equation is solved by adopting a simplified form of the interaction, Eq. (13) in the main text,
where the constants $g_1, g_2, g_3$ may be taken to be the averages of the scattering amplitude (5) over the ranges $0 < p < \Omega$ and $\Omega < p < K$. The solution to the gap equation is then of the form

\begin{align}
\Delta(k) = \Delta_1 \Theta(\Omega - k) + \Delta_2 \Theta(k - \Omega).
\end{align}

Setting $T = 0$ in the gap equation (15) in the main text yields

\begin{align}
\Delta_1 &= - \frac{g_1 \Delta_1 }{2\nu_0} \int_0^{\Omega}{
\frac{ d^2p}{(2\pi)^2}\frac{1}{ \sqrt{\epsilon_p^2 + \Delta_1^2}}
} - \frac{g_2 \Delta_2}{2\nu_0} \int_{\Omega}^\Lambda{
\frac{d^2p}{(2\pi)^2}\frac{1}{ \epsilon_p } 
}, \nonumber \\
\Delta_2 &= - \frac{g_2 \Delta_1}{2\nu_0} \int_0^{\Omega}{ \frac{d^2p}{(2\pi)^2}\frac{1}{ \sqrt{\epsilon_p^2 + \Delta_1^2}}} - \frac{g_3 \Delta_2}{2\nu_0} \int_{\Omega}^{\Lambda}{\frac{d^2p}{(2\pi)^2}\frac{1}{ \epsilon_p} },
\end{align}
where at large $p$, we make the replacement $\sqrt{\epsilon_p^2 + \Delta_2^2} \rightarrow \epsilon_p$, and the single--particle density of states is $\nu_0=k_F/(2\pi v)$. The second line gives

\begin{align}
\Delta_2 = \left( 1 + \frac{g_3}{2\nu_0} \int_{\Omega}^\Lambda{ \frac{d^2p}{\epsilon_p}}\right)^{-1} \left( - \frac{g_2 \Delta_1}{2\nu_0} \int_0^{\Omega}{ \frac{d^2p}{{(2\pi)^2} \sqrt{\epsilon_p^2 + \Delta_1^2}}}\right),\
\end{align}
from which it follows
\begin{align}
 \Delta_1 &= - \frac{g^* \Delta_1 }{2\nu_0}\int_0^{\Omega}{ \frac{d^2p}{ {(2\pi)^2}\sqrt{\epsilon_p^2 + \Delta_1^2}}}
\end{align}
where the effective coupling $g^*$ is 
\begin{align}
g^* &\equiv g_1 - \frac{ \frac{g_2^2}{2\nu_0} \int_{\Omega}^{ \Lambda}{ \frac{d^2p}{{(2\pi)^2}\epsilon_p}}}{ 1 + \frac{g_3}{2\nu_0} \int_{\Omega}^{\Lambda}{ \frac{d^2p}{{(2\pi)^2}\epsilon_p}}} = \frac{g_1 + \frac{g_1 g_3 - g_2^2}{2\nu_0} \int_{\Omega}^{\Lambda}{ \frac{d^2p}{{(2\pi)^2}\epsilon_p}}}{ 1 + \frac{g_3}{2\nu_0} \int_{\Omega}^{\Lambda}{ \frac{d^2p}{{(2\pi)^2} \epsilon_p}}}.
\end{align}

In the weak-coupling limit we may set $g^* \rightarrow g_1$. The magnitude of the coupling $g_1$ (and hence $g^*$) may be varied arbitrarily between the weak and strong coupling regimes by tuning of the density within a narrow range. We may therefore nominally choose $g^*=g_1=\nu_0\Gamma(k_F,k_F)$ since any corrections to this value can be simply compensated by small changes in the density.

Thus the gap equation to leading order in the coupling depends only on $g_1$,

\begin{align}
1 = - \frac{g_1}{2\nu_0} \int_0^{\Omega}{ \frac{1}{ \sqrt{\epsilon_p^2 + \Delta_1^2}} \frac{p dp}{2\pi}} .
\end{align}

Defining $\Delta_1 = v k_F x$, $\Omega=k_F \kappa_0$ and changing variables $p - k_F = k_F \kappa $, we may express the implicit formula for the gap $x = \Delta/E_F$ in the form

\begin{align}
 1 &=  - \frac{g_1 k_F}{4\pi v \nu_0} \int_{-1}^{\kappa_0 - 1}{
\frac{\kappa + 1 }{\sqrt{\kappa^2 + x^2}}  d\kappa
} =  - \frac{g_1}{2} \left(
\sinh^{-1} \frac{\kappa_0 - 1}{x} + \sinh^{-1} \frac{1}{x} + \sqrt{ ( \kappa_0 - 1)^2 + x^2} - \sqrt{1 + x^2}
\right). 
\end{align}
Solution of this implicit equation yields the zero temperature gap.

\section*{Critical Temperature}

In order to obtain the critical temperature, we take the limit $\Delta \rightarrow 0$ in Eq. (15) in the main text which gives

\begin{align}
\Delta(k) = - \frac{1}{2\nu_0} \int_0^K{ \frac{\Gamma^{l=1}( p, k) \Delta(p)} {\epsilon_p} \tanh\frac{\epsilon_p}{2T} \frac{pdp}{2\pi}}.
\end{align}

Similarly to the $T=0$ gap solution, we obtain coupled equations for $\Delta_1, \Delta_2$. Using the same manipulations we find that to leading order we may keep only the first term in the equation for $\Delta_1$. Writing $T = E_F t$ and $p - k_F = k_F \kappa$,

\begin{align}
1 = - \frac{g_1}{2}\int_{-1}^{ k_0 - 1}{ \frac{\tanh \frac{\kappa}{2t}}{ \kappa} ( \kappa + 1) d\kappa}.
\end{align}

The critical temperature $T_c$ is plotted in Figure 3b. We note that the ratio remains $T_c/\Delta \approx 1.8$ for all $0 < \Delta < 0.2 E_F$.

\section*{Symmetry Analysis of the Superconducting Gap}

The structure of the superconducting order parameter can be obtained by a minimizing the free energy obtained from mean field theory. The generic mean field Hamiltonian to account for all pairing possibilities is 
\begin{equation}
\label{Hmf}
{\cal H}_{MF}=\sum_{\bm k, s, \tau}\varepsilon_{\bm k} c^\dag_{\bm k s \tau}c_{\bm k s \tau} +\sum_{\bm k,s, \tau,s', \tau'} c^\dag_{\bm k s \tau} \left(i\Delta_{\bm k}  s^y \tau^x\right)_{s \tau, s' \tau'} c^\dagger_{-\bm k s \tau} + \text{h.c.} + \frac{1}{2}\Delta_{\bm k}^\dagger \Gamma^{-1}(\bm k ,\bm p)\Delta_{\bm p}
\end{equation}
where $s = \uparrow,\downarrow; \tau = \pm 1$ account for physical spin and valley, with $\tau^\mu$ and $s^\mu$ the corresponding Pauli operators, $c^\dagger_{\bm{k} s \tau}$ creates an electron in the upper energy band, and $\Gamma^{-1}(\bm k ,\bm p)$ is the Cooper channel scattering amplitude. Spin and valley indices in $\Gamma^{-1}(\bm k ,\bm p)$ and $\Delta_{\bm k}$ are implied. We write the gap function as  {
\begin{align}
\Delta_{\bm k}&=\sum_{\alpha,\beta =\pm}\sum_{\mu}d^\mu_{\alpha\beta}s^\mu( \hat{k}_x + i\alpha  \hat{k}_y)\left(\tau^{x} + i\beta \tau^{y}\right)
\end{align}

The basis functions $ \hat{k}_x\pm i \hat{k}_y=(k_x\pm i k_y)/k$} and $\tau^{x} \pm i \tau^{y}$ account for $p$-wave pairing within the two valleys; the analysis can be easily generalised to the case of point group rather than full rotational symmetry -- for more detail see e.g. \cite{Scheurer2019}. Intravalley pairing is accounted for by the factor $\tau^x$ in the Hamiltonian \eqref{Hmf}, since the basis $(\tau^{x} \pm i \tau^{y})\tau^x$ is diagonal in valley space. The vector $\mu$ components of $d^\mu_{\alpha\beta}$ account for the triplet spin structure in the usual way,

\begin{align}
d^x=\frac{1}{2}\left(\ket{\uparrow\uparrow}-\ket{\downarrow\downarrow}\right) \ \ \ \ d^y=\frac{1}{2i}\left(\ket{\uparrow\uparrow}+\ket{\downarrow\downarrow}\right) \ \ \ \ d^z=-\frac{1}{2}\left(\ket{\uparrow\downarrow}+\ket{\downarrow\uparrow}\right).
\end{align}

We collect all degrees of freedom into a triplet (indexed by $\mu=1,2,3$) of $2\times2$ matrices, which are spanned by Pauli matrices $\hat\eta^x, \hat\eta^y, \hat\eta^z$, such that 

\begin{align}
 \hat{\bm{d}}&=(\hat{d}^{x}, \hat{d}^{y}, \hat{d}^{z})\ ; \ \ \ 
 \hat{d}^\mu=\begin{pmatrix}d^\mu_{++} && d^\mu_{+-} \\
 d^\mu_{-+} && d^\mu_{--} \end{pmatrix}.
 \end{align}
 The corresponding symmetry operations then take the following representation
 
 \begin{align}
\notag SU(2)\ \ &: \ \ \ {\bm d}_{\alpha\beta} \hspace{0.05cm}  \mapsto  {\mathcal R}(\theta) {\bm d}_{\alpha\beta}\\
\notag  U(1)_v\ \ &: \ \ \ \hat{d}^\mu \ \ \mapsto  \hat{d}^\mu e^{i\frac{1}{2}\phi\hat\eta^z}\\
\notag U(1)\ \ &: \ \ \ \hat{d}^\mu \ \ \mapsto  e^{i\frac{1}{2}\psi \hat\eta^z}\hat{d}^\mu\\
\Theta \ \ &: \ \ \ \hat{d}^\mu \ \ \mapsto \hat\eta^x  (\hat{d}^\mu)^* \hat\eta^x\ .
 \end{align}
 Here ${\mathcal R}(\theta)$ is a three-dimensional rotation matrix, and $\alpha,\beta=\pm$.

The action { for the mean field Hamiltonian} is given by { 
\begin{align}
{\cal S}&=\frac{1}{2}\sum_{\bm k} \begin{pmatrix}{c}_{\bm k}^\dag \\ {c}_{-\bm k} \end{pmatrix}^T \begin{pmatrix}\omega- \varepsilon_{\bm k}  & -\Delta_{\bm k}i s^y \tau^x \\ i s^y \tau^x\Delta_{\bm k}^\dag & \omega + \varepsilon_{\bm k} \end{pmatrix} \begin{pmatrix}{c}_{\bm k} \\ {c}_{-\bm k}^\dag \end{pmatrix}-\frac{1}{2}\Delta_{\bm k}^\dagger \Gamma^{-1}(\bm k ,\bm p)\Delta_{\bm p}.
\end{align}
}Integrating out the electronic degrees of freedom gives the free energy in terms of the order parameter $d^\mu_{\alpha\beta}$ \cite{Kleinert1978,Sigrist2005}, 

{ 
\begin{align}
{\cal S}&=-\frac{i}{2}\text{Tr} \log \begin{pmatrix}\omega- \varepsilon_{\bm k}  & -\Delta_{\bm k}i s^y \tau^x \\ i s^y \tau^x\Delta_{\bm k}^\dag & \omega + \varepsilon_{\bm k} \end{pmatrix} -\frac{1}{2}\Delta_{\bm k}^\dagger \Gamma^{-1}(\bm k ,\bm p)\Delta_{\bm p} \equiv i\mathcal F^{(0)} + i\mathcal F
\end{align}
where $\mathcal F^{(0)}$ is the free energy for free fermions and $\mathcal F$ is given by 
\begin{align}
\mathcal F = \frac{1}{2}\sum_{n=0}^\infty \frac{(-1)^{2n}}{2n}\text{Tr}\left(  \frac{i}{\omega- \varepsilon_{\bm k}} \Delta_{\bm k} \frac{i}{\omega+ \varepsilon_{\bm k}}\Delta_{\bm k}^\dagger \right)^{2n} -\frac{1}{2}\Delta_{\bm k}^\dagger \Gamma^{-1}(\bm k ,\bm p)\Delta_{\bm p} 
\end{align}

In the weak coupling limit, we are justified in approximating $\mathcal F$ by the quartic and quadratic terms in this series, which gives us a Landau-Ginzburg free energy of the form

\begin{align}
\label{FE}
\notag{\cal F}&={\cal F}^{(2)} + {\cal F}^{(4)}\\
\notag{\cal F}^{(2)} &= -a_1\left(\bm{d}^\dag_{++}\bm{d}_{++}+\bm{d}^\dag_{--}\bm{d}_{--}\right)+a_2\left(\bm{d}^\dag_{+-}\bm{d}_{+-}+\bm{d}^\dag_{-+}\bm{d}_{-+}\right)\\
\notag{\cal F}^{(4)}&=\sum_{\beta}\Big\{\frac{1}{2}b\left(\bm{d}^\dag_{+\beta}\bm{d}_{+\beta} \right)^2+\frac{1}{2}b\left(\bm{d}^\dag_{-\beta}\bm{d}_{-\beta} \right)^2 +2b\left(\bm{d}^\dag_{+\beta}\bm{d}_{+\beta} \right)\left(\bm{d}^\dag_{-\beta}\bm{d}_{-\beta} \right)\\
 & \ \ \ \ \ + 2b \ \left( \bm{d}^*_{+\beta}\times\bm{d}_{+\beta}\right)\left( \bm{d}^*_{-\beta}\times\bm{d}_{-\beta}\right) + \frac{1}{2}b \sum_\alpha|\bm{d}^*_{\alpha\beta}\times\bm{d}_{\alpha\beta}|^2\Big\}
\end{align} }
where $a_1, a_2>0$. Importantly, the coefficients of the quartic terms are all known and are simple ratios of each other, and $b$ is positive, {
\begin{align}
b \propto \sum_{\omega, {\bm k}}\left[\frac{1}{(\omega^2 - \varepsilon^2_{\bm k})^2}\right]>0.
\end{align}  }

\begin{figure}[t]
	\includegraphics[width = 0.3\textwidth]{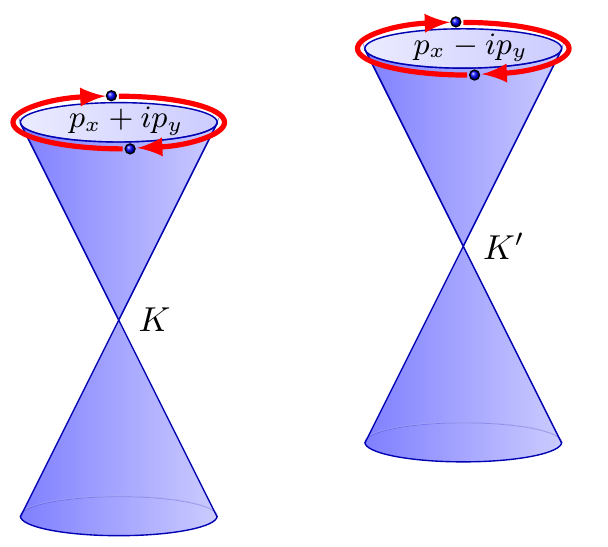}
	\caption{ In each valley, electrons undergo pairing with angular momentum $|\ell|=1$. The lowest energy superconducting state \eqref{suppsymmetry} is a time reversal invariant combination of opposite chirality states at each Dirac point. The electrons undergo $p_x+ip_y$ pairing in one valley, and $p_x-ip_y$ pairing in the other.}
	\label{fig:symmetry}
\end{figure}

The off--diagonal terms in $\bm d_{\alpha\beta}$ with $\alpha=-\beta$ are energetically costly and can be set to zero. Since $\alpha=\beta$, this implies that condensation within each valley involves a single chirality. The cross products make non-unitary states energetically unfavorable. We find that the free energy for $\bm d_{++}$ and $\bm d_{--}$ decouple.  
Minimizing (\ref{FE}), we find that 
\begin{align}
\notag \bm d_{++}&= {\cal R} d_0(1,0,0), \hspace{0.5cm} \bm d_{-+}= \bm 0, \hspace{2.35cm} d_0^2 = \frac{a_1}{b}\\
\bm d_{-+}&= \bm 0, \hspace{2.35cm}  \bm d_{--}= {\cal R} d_0(1,0,0), \hspace{0.5cm} d_0^2 = \frac{a_1}{b}
\end{align}
where ${\cal R}$ is an arbitrary $SO(3)$ rotation matrix. Setting $\bm d_{++}=\bm d_{--}\equiv {\bm d}/ 2$, the gap function then takes the form
\begin{align}
\notag \Delta_{\bm k}is^y\tau_x &= \sum_{\alpha \pm}\sum_{\mu}d^\mu_{\alpha\beta}s^\mu is^y  ( \hat{k}_x + i\alpha \hat{k}_y )\left(\tau^{0} + \alpha \tau^{z}\right)\\
&=({\bm d}\cdot{\bm s}\  is^y)( \hat{k}_x \tau^0 + i\hat{k}_y \tau^z )
\label{suppsymmetry}
\end{align}

This is a unitary spin triplet state, with chiral $p_x+ip_y$ pairing in one valley and $p_x-ip_y$ pairing in the other -- analogous to two (opposite chirality) copies of the A--phase of $^3$He \cite{Leggett1975}. This state preserves time reversal symmetry, but spontaneously breaks $SU(2)$ spin rotation symmetry down to $U(1)$ rotations about the vector $\bm d$ (which is arbitrary).

\end{document}